\documentclass[journal]{IEEEtran}
\usepackage[english]{babel}
\usepackage[utf8]{inputenc}
\usepackage{amsmath}
\usepackage{amssymb}
\usepackage{bbm}

\usepackage{graphicx}
\usepackage{amstext}
\usepackage{amsmath}
\usepackage{amsfonts,amssymb}
\usepackage{amsmath,tabu}
\usepackage{amssymb}
\usepackage[dvips]{epsfig}
\usepackage{epsfig}
\usepackage[dvips]{graphicx}
\usepackage{multirow}
\usepackage[utf8]{inputenc}
\usepackage[english]{babel}
\usepackage[linesnumbered,ruled]{algorithm2e}
\usepackage{float}
\usepackage[nospace,noadjust]{cite}
\usepackage{bm}
\usepackage{xcolor,colortbl}
\usepackage{subcaption}

\newcommand{\rank}{\textrm{rank}}
\newcommand{\srank}{\textrm{rank}_{S}}
\providecommand{\keywords}[1]{\textbf{\textit{Index terms---}} #1}
\makeatletter
\def\thanks#1{\protected@xdef\@thanks{\@thanks
		\protect\footnotetext{#1}}}
\makeatother

\usepackage{color}

\allowdisplaybreaks[4]

\definecolor{Gray}{gray}{0.85}
\definecolor{White}{gray}{1}
\newcolumntype{a}{>{\columncolor{Gray}}c}
\usepackage{amsmath}
\usepackage{kantlipsum}

\usepackage{siunitx}

\newcommand{\bfnum}[1]{\text{\bfseries{\num{#1}}}}

\begin{document}
	\title{{Exploiting Cognition in} ISAR Processing for Spectral Compatibility Applications}
	
	\author{ Massimo Rosamilia,~\IEEEmembership{Member,~IEEE}, Augusto Aubry,~\IEEEmembership{Senior Member,~IEEE}, Alessio Balleri,~\IEEEmembership{Senior Member,~IEEE}, Antonio De Maio,~\IEEEmembership{Fellow,~IEEE}, and Marco Martorella,~\IEEEmembership{Fellow,~IEEE}
		\thanks{The work of Massimo Rosamilia, Augusto Aubry, and Antonio De Maio was partially supported by the European Union under the Italian National Recovery and Resilience Plan (NRRP) of NextGenerationEU, partnership on ``Telecommunications of the Future'' (PE00000001 - program ``RESTART'').}
		\thanks{M. Rosamilia, A. Aubry, and A. De Maio are with the Department of Electrical Engineering and Information Technology, Università degli Studi di Napoli ``Federico II”, DIETI, Via Claudio 21, I-80125 Napoli, Italy and with the National Inter-University Consortium for Telecommunications, 43124 Parma, Italy (E-mail: massimo.rosamilia@unina.it, augusto.aubry@unina.it, ademaio@unina.it). (Corresponding author: Antonio De Maio.)} 
		\thanks{A. Balleri is with the Centre for Defence Engineering and Applied Science, Cranfield University, Defence Academy of the United Kingdom, Shrivenham, SN6 8LA, United Kingdom (E-mail: a.balleri@cranfield.ac.uk).}
		\thanks{Marco Martorella is with the Department of Electronic, Electrical and Systems Engineering, University of Birmingham, B15 2TT Birmingham, U.K., and also with the Radar and Surveillance Systems (RaSS) National Laboratory of CNIT, 56124 Pisa, Italy (e-mail: marco.martorella@cnit.it).
		}
		\thanks{This paper was presented in part at the 2024 IEEE International Workshop on Technologies for Defense and Security (TechDefense)~\cite{10863447}.
		}
	}
	
	\markboth{}%
	{Shell \MakeLowercase{\textit{et al.}}: Bare Demo of IEEEtran.cls for Journals}
	\maketitle
	
	\begin{abstract}
		This paper introduces and analyzes the concept of a cognitive inverse synthetic aperture radar (ISAR) ensuring spectral compatibility in crowded electromagnetic environments. In such a context, the proposed approach alternates between environmental perception, recognizing possible emitters in its frequency range, and an action stage, synthesizing and transmitting a tailored radar waveform to achieve the desired imaging task while guaranteeing spectral coexistence with overlaid emitters. The perception is carried out by a spectrum sensing module providing the true relevant spectral parameters of the sources in the environment. The action stage employs a tailored signal design process, synthesizing a radar waveform with bespoke spectral notches, enabling ISAR imaging over a wide spectral bandwidth without interfering with the other radio frequency (RF) sources. A key enabling requirement for the proposed application is the capability to successfully recover possible missing data in the frequency domain (induced by spectral notches) and in the slow-time dimension (enabling concurrent RF activities still in a cognitive fashion). This process is carried out by resorting to advanced methods based on either the compressed-sensing framework or a rank-minimization recovery strategy. The capabilities of the proposed system are assessed exploiting a dataset of drone measurements in the frequency band between 13 GHz and 15 GHz. Results highlight the effectiveness of the devised architecture to enable spectral compatibility while delivering high-quality ISAR images as well as additional RF activities.
		
	\end{abstract}

	\section{Introduction}
	The electromagnetic spectrum is a valuable and highly dynamic resource that is {yet to be fully harnessed by radar systems}~\cite{7838312}. Accordingly, an essential requirement for a next-generation radar system is the ability to {operate} in a congested electromagnetic {spectral environment and} guarantee compatibility with overlaid emitters {by agility} in the exploitation of the dynamically available frequency bands.
	Recent research emphasizes the importance of the waveform design and diversity (WDD) paradigm~\cite{6081358, 1245056, wicks2006waveform, gini2012waveform,iet:/content/books/ra/sbra533e} as a means of addressing this challenge. This approach, when implemented within a cognitive radar architecture, involves alternating between two key steps: a perception stage, where the radar detects emitters within its band of interest~{\cite{8574991, 9187976}}, and an action stage, where a spectrally-shaped waveform is designed and transmitted to minimize interference with {the aforementioned sources} and {to} enhance spectrum efficiency by optimizing the signal-to-interference-plus-noise ratio (SINR)~\cite{5604089, 5960706, 6586024, 7225102, 7414411, 7838312, farina2017impact, 9052442}.

	While the open literature has mainly focused on radar detection, the present study shifts the focus to the radar imaging. Specifically, this work proposes a cognitive inverse synthetic aperture radar (ISAR) imaging system with spectral compatibility requirements. {As part of the radar operations workflow, a perception mode provides awareness of the emitters in the surrounding environment via specific radio frequency (RF) sensing modules~\cite{8574991, 9187976, 10891687}. This allows} the synthesis and transmission of a {tailored} radar waveform encompassing spectral notches to avoid {interference} with overlaid systems, {realizing the so-called \textit{perception-action cycle}}. Consequently, the radar is able to perform ISAR imaging over a wide bandwidth without {inducing} {and experiencing} interference from coexisting emitters.

	Nevertheless, the presence of spectral notches in the transmitted waveform inevitably results in frequency domain gaps, which may eventually lead to spurious and high sidelobes in the resulting ISAR image if not adequately accounted for. {Furthermore, in the context of a multifunction phased array radar (MPAR) system, wherein multiple tasks compete for the limited time slots, a tailored strategy to maximize the available resources {(according to scheduling policies~\cite{10510313})} might assign non-contiguous or temporally uneven pulses to the imaging of a target, resulting in some slow-time domain gaps in the collected data.}
	To overcome such drawbacks, a {data recovery process is demanded} at the image formation stage in order to successfully accomplish ISAR imaging even in the presence of missing data in the frequency/slow-time domain~\cite{7272859, tomei2016compressive, bacci2016compressive, ghaffari2009sparse}. In the past decade, the compressed sensing framework has shown to be remarkably powerful in tackling this kind of problem, with numerous algorithms proposed in the open literature~\cite{blumensath2009iterative, needell2009cosamp, 4385788, 7979584}. In a nutshell, this strategy exploits the inherent property of the collected data exhibiting a sparse representation in some basis (e.g., Fourier)~\cite{1614066}. Notably, recent recovery algorithms leveraged on structures beyond {conventional} sparsity, such as group sparsity, low-rankness, etc., {to regularize the inference process}~\cite{7979584, fazel2008compressed, recht2010guaranteed, peyre2011group, liu2014exact}.
		
	For the problem at hand, the well-known 2D-SL0 algorithm~\cite{ghaffari2009sparse, 6581073, tomei2016compressive} is applied to retrieve the missing data in the considered ISAR application. Furthermore, a low-rank recovery process is formulated to reconstruct the data based on the assumption of a limited amount of dominant modes in the image. {After} approximating the rank with its tightest convex description, the problem is optimally solved by resorting to the iterative Majorization Minimization (MM) algorithm~\cite{wu2010mm, hunter2004tutorial, sun2016majorization}.

	To assess the performance of the proposed cognitive ISAR system, {an} extensive numerical {analysis has} been conducted employing a dataset of drone measurements in the frequency band $[13, 15]$ GHz and HH polarization~\cite{9982651, 10226589}. The results highlight the radar capability to operate within a congested spectrum while producing high-quality ISAR images.
	
	The remainder of this work is organized as follows. Section II introduces the waveform design framework for spectral compatibility applications. {Section III discusses the ISAR imaging technique in the presence of incomplete data in the frequency/slow-time domain by presenting two recovery strategies based on the CS framework and rank-minimization, respectively.} In Section IV, extensive numerical analysis on a dataset comprising returns from a {commercial} drone is reported and discussed considering several scenarios of practical relevance. Finally, Section V concludes the paper.

	\subsection{Notation}
	Boldface is used for vectors $\bm{a}$ (lower case), and matrices $\bm{A}$ (upper case).  The $n$th element of $\bm{a}$ and the $(m,l)$th entry of $\bm{A}$ are, respectively, denoted by $\bm{a}(n)$ and $\bm{A}(m,l)$. The transpose, the conjugate, and the conjugate transpose operators are denoted by the symbols $(\cdot)^\mathrm{T}$, $(\cdot)^\star$ and $(\cdot)^\dagger$, respectively. {${\mathbb{C}}^N$ is the set of $N$-dimensional vectors of complex numbers.} {Furthermore, $\bm{1}$ is the vector with all elements being one.} For any complex number $x$, $|x|$ indicates the modulus of $x$. For any matrix $\bm{A}$, $\|\bm{A}\|_0$, {$\|\bm{A}\|_1$},  $\|\bm{A}\|_F$, and $\|\bm{A}\|_\star$ denote the corresponding $\ell_0$, {$\ell_1$}, Frobenius, nuclear norm, respectively, whereas $\left(\bm{A}\right)^+$ represents its pseudoinverse. In addition, {$\|\bm{A}\|_{\textrm{max}}=\max_{m,l} |\bm{A}(m,l)|$ is the max norm of $\bm{A}$.} {Moreover, for any real matrix $\bm{B}$,  $\left(\bm{B}\right)_+$ denotes the rectified linear unit operator, which replaces all negative entries of $\bm{B}$ with zero.} {For any real number $a$, $\left\lceil a \right\rceil$ yields the least integer greater than or equal to $a$.} In addition, the {rank} and the Shannon-rank of $\bm{A}$ are denoted by $\rank\{A\}$ and $\srank\{A\}$, respectively. Finally, $\mathbb{A}[\cdot]$ represents the spatial mean operator, whereas the letter $j$ represents the imaginary unit (i.e., $j=\sqrt{-1}$).

	\section{Waveform design for spectral compatibility}
	In this section, the problem of designing spectrally-shaped {waveforms} is briefly introduced with reference to {a} {cognitive radar system}.
	
	{To this end,} let us consider a radar operating in the presence of $K$ {emitters}, each of them transmitting over a frequency band  $\Omega_k = [f_1^k, f_2^k], k=1,\dots, K$, within the {radar} frequency {support} range, where $f_1^k$ and $f_2^k$ denote the lower and upper normalized frequencies (with respect to the sampling {rate}) for the $k$th {source}, respectively.
	
	Let $c(t)$ be the baseband equivalent of the radar transmitted waveform whose complex sequence $\bm{c}$ ({corresponding to samples of $c(t)$ picked up at} the frequency rate equal to the {radar} bandwidth) is $c(i), \; i=1, \dots, N$.
	To ensure spectral compatibility with the overlaid emitters, the cognitive radar has to suitably shape {the} spectrum {of the transmitted signal} to manage the amount of interfering energy produced in the shared frequency {intervals}. {The average} {signal} energy transmitted on the $k$th band $\Omega_k$ is given by~\cite{7414411}
	\begin{equation}\label{eq:avg_energy}
		\frac{1}{f_2^k - f_1^k} \int_{f_1^k}^{f_2^k} S_c(f) df,
	\end{equation}
	where
	\begin{equation}
		S_c(f) = \left|\sum_{n=1}^{N} c(n) e^{-j 2 \pi f (n-1)}\right|^2 = N |\bm{c}^\dagger \bm{p}_f|^2,
	\end{equation}
	with
	\begin{equation}
		{\bm{p}}_{{f}} = \frac{1}{\sqrt{N}} [1, \exp(-j2\pi {{f}}), \dots, \exp(-j2\pi(N-1) {{f}})]^\mathrm{T}
	\end{equation}
	the {temporal steering vector tuned to the} {normalized} frequency ${{f}}$.
	That said, each frequency band $\Omega_k$ is {substantially} discretized by taking the normalized frequencies {$f_i = i/N, \; i=0, \dots, N-1$} {falling in $\Omega_k$ itself}. Then, denoting by
	\begin{equation}\label{eq:fourier_mat} 
		\bm{F}_N(m,n) = \frac{1}{\sqrt{N}} e^{-\frac{j 2 \pi (m-1)(n-1)}{N}}
	\end{equation}
	the $N \times N$ Fourier matrix scaled {by} $1/\sqrt{N}$, a {viable means} to quantify the average energy {level} {injected} on $\Omega_k$ is {through}
	\begin{equation}
		\frac{N {\Delta f}}{f_2^k - f_1^k} \sum_{{f_i \in \Omega_k}} |\bm{c}^\dagger \bm{p}_{f_i}|^2 = \frac{1}{f_2^k - f_1^k} \sum_{{f_i \in \Omega_k}} |\bm{c}^\dagger \bm{p}_{f_i}|^2 = \bm{c}^\dagger{\bm{R}_N^k}\bm{c},
	\end{equation}
	where\footnote{{Unless otherwise specified, the subscript $N$ in $\bm{F}_N$, $\bm{Q}_N$, and $\bm{R}_N^k$ denotes the number of rows and represents the number of points used to discretize {the frequency interval $[0,1]$}.}}
	\begin{equation}
		{\bm{R}_N^k} =\frac{1}{f_2^k - f_1^k} {\bm{Q}_N^k} {\bm{Q}_N^k}^\dagger,
	\end{equation}
	and ${\bm{Q}_N^k}$ is the submatrix of $\bm{F}_N$ whose columns correspond to the {discrete} frequencies in the interval $\Omega_k$.

	Thus, {indicating} by $E_I^k, \; k=1,\ldots,K$, the acceptable level of {interference} on {$\Omega_k$}, which is related to the quality of service {demanded} by the $k$th {{licensed} emitter}, the transmitted {radar} waveform has to comply with the constraints
	\begin{equation}\label{eq:spectral_constr}
		\bm{c}^\dagger{\bm{R}_N^k}\bm{c}\leq E_I^k,\quad k=1,\ldots,K .
	\end{equation}
	By doing so, a detailed control of the interference energy produced on each shared frequency bandwidth is enforced.

In the following, a waveform design algorithm for spectral compatibility applications is presented, which {complies with} the aforementioned spectral constraints.

\subsection{QCQP-based block waveform design}
As previously mentioned, the purpose of the action stage in the considered cognitive radar architecture is to synthesize a bespoke radar waveform in order to control the amount of transmitted energy on overlaid transmitters. To this end, it is assumed that at the perception stage, the cognitive radar gathers {true} spectral knowledge on the sources operating in its frequency band through a spectrum sensing module that periodically monitors the RF environment~\cite{8574991, 10891687}. 

Leveraging the information from the perception stage, a bespoke {signal} is generated using a waveform design algorithm and is transmitted via the radar antenna towards the target location. A modified version of the algorithm derived in~\cite{7414411} is {now proposed}, which {relies} on a quadratically constrained quadratic program (QCQP) formulation attempting to control the amount of interfering energy produced in the frequency intervals occupied by overlaid EM sources, while maximizing the similarity with a reference signal.

{Precisely}, the similarity with a standard chirp signal $\bm{c}_0$  (tailored for the ISAR requirements in terms of spectral bandwidth) is considered as figure of merit, in order to perform the spectral shaping {of} the cognitive waveform.
{Hence}, by partitioning the waveform to synthesize in $L$ smaller blocks of size $\bar{N}$ (assuming, without loss of generality, that $N = L \bar{N}$), i.e.,
\begin{equation}
	\bm{c} = [\bm{c}_1^\mathrm{T}, \dots, \bm{c}_{L}^\mathrm{T}]^\mathrm{T},
\end{equation}
the design approach {is tantamount to minimizing the distance $\|\bm{c}-\bm{c}_0 \|^2$ and} can be formulated as the following QCQP convex optimization problem
\begin{equation}\label{eq:qcqp_whole}
	\mathcal{P} \begin{cases}{\min\limits_{\bm{c}\in \mathbb{C}^{{N}}}} & \|\bm{c}-\bm{c}_0 \|^2\\ \mbox{s.t.} & 
		{\Vert {\bm{c}_l}\Vert ^2\leq 1/L}, \; l=1,\dots, L \\ &
		{\bm{c}}^{\dagger}{{\bm{R}_N^k}}{\bm{c}}\leq E_I^k,\,k=1,\ldots K\end{cases},
\end{equation}
whose optimal solution can be obtained in polynomial time with arbitrary precision.
{The norm constraint imposed on the individual blocks ${\bm{c}_i}$ of the sequence $\bm{c}$ is aimed at ensuring a bound on the transmitted energy, along with the avoidance of possible energy imbalances {of different segments in} the time domain.}

In practical {wideband} scenario where the waveform length $N$ is much larger than $10^{{4}}$, computing a solution to $\mathcal{P}$ involves handling a rather high computational and space complexity (i.e., demanding the storage of large matrices). In such instances, this drawback can be {potentially} addressed by sequentially optimizing each block $\bm{c}_l, \; l=1, \dots, L$, leading to a (generally) sub-optimal and {computationally} tractable design process.
By partitioning the reference code $\bm{c}_0$ in $L$ blocks of length $\bar{N}$, i.e.,
\begin{equation}
	\bm{c}_0 = [\bm{c}_{0,1}^\mathrm{T}, \dots, \bm{c}_{0,l}^\mathrm{T}, \dots, \bm{c}_{0,L}^\mathrm{T}]^\mathrm{T},
\end{equation}
the design of the $l$th sub-block involves the maximization of the similarity with the reference signal {segment} ${\bm{c}_{0,l}}$ while satisfying tailored spectral constraints for the considered block, i.e.,
{\begin{equation}\label{eq:spectral_constr2}
		\bm{c}_l^\dagger \bm{R}_{\bar{N}}^k\bm{c}_l\leq E_I^k/L, \quad k=1,\ldots,K .
\end{equation}}
Nevertheless, {synthesizing} each block {$\bm{c}_l$} independently leads to a lack of control {of} the spectral {behaviour} between consecutive blocks. {To guarantee a smooth transition between the subsequent blocks of samples and avoid {possible} spectral spuries, an heuristic but  effective solution technique is now illustrated. Specifically, the sequence $\bm{c}$ is progressively designed by employing a set of {partially overlapped} time windows (i.e., blocks) each including a segment of the signal already synthesized and a segment to be optimized.} {{Accordingly}, denoting by $\bm{c}_{l-1,W}$ the vector composed of the last $W \in [0, \bar{N}/2]$ elements of $\bm{c}_{l-1}$, i.e., the  signal block already designed, the optimization process reformulates the quadratic constraints in terms of $\bm{\tilde{c}} =[{\bm{c}}_{l-1,W}^\mathrm{T}, {\bm{\check{c}}_l}^\mathrm{T}]^\mathrm{T} \in \mathbb{C}^{\bar{N}}$, with $\bm{\check{c}}_l$ the segment to optimize {at the current step} (of size $\bar{N}-W$).}

Moreover, the total number of optimizations to perform (in addition to the one pertaining to the first block) is equal to $\tilde{L} = \left\lceil (N-\bar{N})/(\bar{N}-W) \right\rceil$.

\begin{algorithm}[t]
	\caption{QCQP block waveform design for cognitive radar}\label{alg:qcqp}
	\KwIn{$N$, $\bar{N}$, $W$, $\{{\bm{R}_{\bar{N}}^k}\}_{k=1}^K$, $\{E_I^k\}_{k=1}^K$, {$\bm{c}_0$}.}
	set $L = \left\lceil N/\bar{N} \right\rceil$ and $\tilde{L} = \left\lceil (N-\bar{N})/(\bar{N}-W) \right\rceil$\;
	{partition $\bm{c}_0$ in ${\tilde{L}+1}$ blocks as $\bm{c}_0 = [\bm{\bar{c}}_{0,1}^\mathrm{T}, \bm{\bar{c}}_{0,2}^\mathrm{T}, \dots, \bm{\bar{c}}_{0,\tilde{L}+1}^\mathrm{T}]^\mathrm{T}$, with $\bm{\bar{c}}_{0,1}$ of size $\bar{N}$ and ${\bm{\bar{c}}_{0,l}}$ of size $(\bar{N}-W), \, l=2, \dots, {\tilde{L}+1}$}\;
	compute $\bm{c}_1$ as a solution to~\eqref{eq:qcqp_first}\;
	\For{$l=[2, \dots, {\tilde{L}+1}]$}{
		compute $\bm{c}_l$ as a solution to~\eqref{eq:qcqp_block};
	}
	\KwOut{$\bm{c} = [\bm{c}_1^\mathrm{T}, \dots, \bm{c}_{{\tilde{L}+1}}^\mathrm{T}]^\mathrm{T}$}
\end{algorithm}

By leveraging the aforementioned guidelines, the waveform design problem for the first and the $l$th ({with $l=[2, \dots, \tilde{L}+1]$}) block can be formulated as the following QCQP convex optimization problems
{\begin{equation}\label{eq:qcqp_first}
		\mathcal{P}_1 \begin{cases}{\min\limits_{\bm{\check{c}}_1\in \mathbb{C}^{\bar{N}}}} & \|\bm{\check{c}}_1-\bm{\bar{c}}_{0,1} \|^2\\ \mbox{s.t.} & 
			\Vert {\bm{\check{c}}_1}\Vert ^2\leq 1/L \\ &	
			{\bm{\check{c}}_1}^{\dagger}{{\bm{R}_{\bar{N}}^k}}{\bm{\check{c}}_1}\leq E_I^k / L,\,k=1,\ldots K\end{cases}
	\end{equation}
	and
	\begin{equation}\label{eq:qcqp_block}
		\mathcal{P}_l \begin{cases}{\min\limits_{{\bm{\check{c}}_l}\in \mathbb{C}^{\bar{N}{-W}}}} & \|\bm{\check{c}}_l-{\bm{\bar{c}}_{0,l}} \|^2\\ \mbox{s.t.} & 
			\bm{\tilde{c}} = [{\bm{c}}_{l-1,W}^\mathrm{T}, {\bm{\check{c}}_l}^\mathrm{T}]^\mathrm{T} \\ &		\Vert {\bm{\tilde{c}}}\Vert ^2 \leq 1/{L}\\ &
			{\bm{\tilde{c}}}^{\dagger}{{\bm{R}_{\bar{N}}^k}}{\bm{\tilde{c}}}\leq E_I^k/{L},\,k=1,\ldots K\end{cases}
	\end{equation}
	respectively, where the reference code $\bm{c}_0$ is partitioned in ${\tilde{L}+1}$ blocks as
	\begin{equation}
		\bm{c}_0 = [\bm{\bar{c}}_{0,1}^\mathrm{T}, \bm{\bar{c}}_{0,2}^\mathrm{T}, \dots, \bm{\bar{c}}_{0,\tilde{L}+1}^\mathrm{T}]^\mathrm{T},
	\end{equation}
	with $\bm{\bar{c}}_{0,1}$ of size $\bar{N}$ and ${\bm{\bar{c}}_{0,l}}$ of size $(\bar{N}-W), \, l=2, \dots, {\tilde{L}+1}$.}
Remarkably,~\eqref{eq:qcqp_first} and~\eqref{eq:qcqp_block} can be {practically handled} {as they are} computationally and space efficient.

{Summarizing,} by combining block-wise optimization with different {segments of the} reference sequence and {tailored} spectral constraints, the proposed approach {exhibits} an affordable computational effort {while satisfying the spectral requirements and effectively mimicking the characteristics of the reference waveform}.
\textbf{Algorithm~\ref{alg:qcqp}} {specifies} the proposed block-wise waveform design strategy for the considered cognitive ISAR system operating to guarantee spectral coexistence with {other} emitters.

\subsection{{Analysis of the Cognitive-based Spectral Compatible ISAR Waveform}}
In the following, a waveform synthesized with \textbf{Algorithm~\ref{alg:qcqp}} is analyzed in terms of spectral {features} and autocorrelation behaviour. At the synthesis stage, a radar bandwidth of $2$ GHz and a waveform duration of 5 $\mu$s, are considered as system parameters, with a standard chirp used as reference signal $\bm{c}_0$. {The synthesis of the spectral compatible waveform is performed via \textbf{Algorithm~\ref{alg:qcqp}} using block size $\bar{N}=5000$ and $W=2500$.} For the cognitive application, {two} {licensed} emitters are assumed operating in the environment, on the intervals $[13.38, 13.62]$ GHz and $[14.53, 14.65]$ GHz respectively. Consequently, stop-bands have been employed for spectral shaping, with a notch depth of 40 dB and 30 dB, respectively.

\begin{figure}[t]
	\centering
	\includegraphics[trim=85 10 90 40,clip,width=0.98\linewidth]{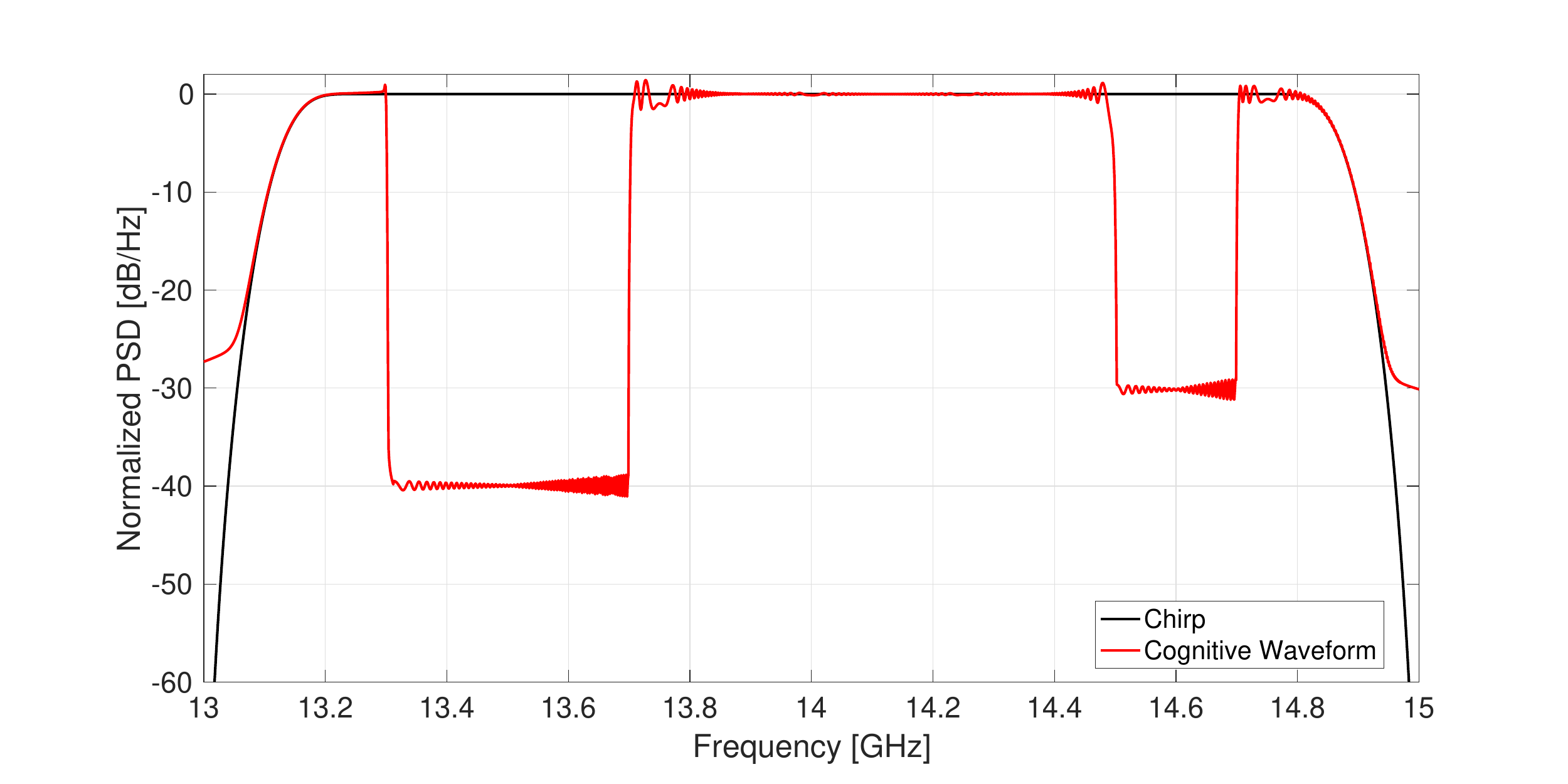}
	\caption{{PSD estimate} of the reference (chirp) waveform and the synthesized one via \textbf{Algorithm~\ref{alg:qcqp}} (using block size {$\bar{N}=5000$}). The PSD are estimated using the Welch method (considering segments of 3000 samples with 2900 samples of overlap from segment to segment, weighted with an Blackman-Harris window and normalized to the mean value of the reference waveform PSD.}
	\label{fig:periodogram_sim} 
\end{figure}

Fig.~\ref{fig:periodogram_sim} reports the power spectral density (PSD) of the reference waveform $\bm{c}_0$ and the cognitive one synthesized using \textbf{Algorithm~\ref{alg:qcqp}}. The PSD are estimated using the Welch method considering segments of 3000 samples, with 2900 samples of overlap from segment to segment, weighted with an Blackman-Harris window. Moreover, the PSDs are normalized to the maximum PSD value of the reference waveform. The {figure} clearly shows that the designed waveform exhibits spectral notches {almost} perfectly lined-up within the desired stop-bands, while still adhering to the reference waveform on the passband intervals. Moreover, {the} figure highlights that the resulting {notches} depth is approximately $-40$ dB and $-30$ dB, with some (limited) spurious near the boundaries of the stop-bands.

\begin{figure}[t]
	\centering
	\includegraphics[trim=45 0 50 10,clip,width=0.95\linewidth]{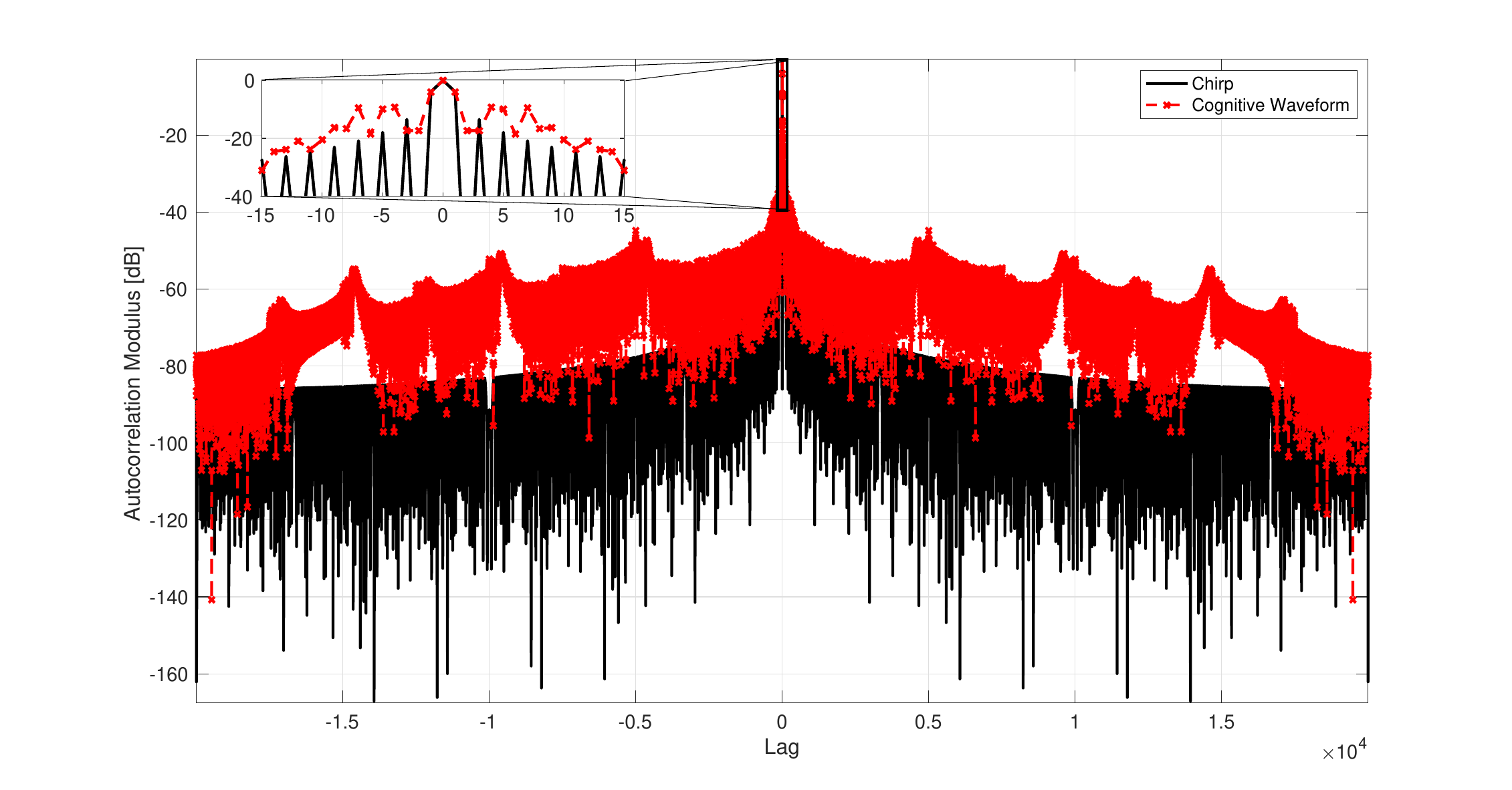}
	\caption{Normalized autocorrelation of the reference and the synthesized waveforms.}
	\label{fig:autocorr_sim} 
\end{figure}

To quantify the characteristics of the designed spectrally-notched radar signal, Fig.~\ref{fig:autocorr_sim} {compares} the {amplitude} of its autocorrelation function (AF) (normalized to {the peak}) with that of the reference signal. The results pinpoint that the obtained radar sequence is characterized by an autocorrelation peak sidelobe level (PSL) of $-9.5$ dB, while for the reference waveform it is equal to $-13$ dB, with an actual offset of about $3.5$ dB. Furthermore, the designed waveform successfully achieves the desired spectral shaping while maintaining the AF 3dB mainlobe {almost} unaltered, matching its reference counterpart.

	\section{ISAR imaging in the presence of missing data}
	ISAR imaging is a technique used to generate high-resolution images of moving targets. Conventional imaging {techniques} rely on the coherent integration of received echoes over a range of aspect angles, generally resorting to Fourier-based methods (downstream a careful compensation stage). However, in the presence of missing samples in the frequency domain, for instance due to spectral notches in the transmitted waveform, ISAR images obtained with the standard range–Doppler (RD) algorithm~\cite{chen2014inverse} may be distorted or exhibit a coarse {final image} resolution due to {the degradation of the matched filter response}. Therefore, to provide high-quality images in this context, the processing stage {could} include a strategy to retrieve any missing {measurements within the available} collected data.
	In particular, let $\bm{S}$ be\footnote{{Note that a tailored pre-processing step {(also accounting for the presence of possible missing data)} is assumed to be carried out to compensate for the translational component in the measured signals~\cite{giusti2013autofocus,martorella2005contrast}.}} the observed \textit{incomplete} {matrix} in the slow-time/frequency domain {(where zeros are present in locations {where} either a frequency or a pulse is missed)} and $\bm{I}$ the ISAR image{; then, the} goal is to minimize the displacement term {(quantified via {the} Frobenius norm)}, i.e.,
	\begin{equation}\label{eq:problem1}
		\min_{\bm{I}} \|\bm{S} - \bm{\Theta}_x \bm{I} \bm{\Theta}_y^\dagger \|_F^2 ,	
	\end{equation}
where $\bm{\Theta}_x$ and $\bm{\Theta}_y$ are the corresponding undercomplete Fourier matrices (obtained by zeroing specific rows of the DFT matrix corresponding to the missing frequencies/pulses {samples})~\cite{tomei2016compressive}, which can be computed starting from the parameters of the cognitive waveform spectral notches and {the location of the {temporal gaps}}.
{Unfortunately}, Problem~\eqref{eq:problem1} is ill-posed since the missing entries of $\bm{S}$ cannot be uniquely determined without making assumptions on the matrix $\bm{I}$, {namely multiple solutions to Problem~\eqref{eq:problem1}, each producing a different filling of the missed measurements, exist.}
{A viable {method} to overcome this {limitation} is to introduce another  figure of merit to assess the quality of the different feasible solutions, which {is usually referred to as regularization process in the open literature {\cite{zare2020determination, bauer2011comparingparameter, 9521836}} and} amounts to substantially limit the solution state space.} {Otherwise stated the inference task can be reformulated as} a bi-objective {optimization problem}, {where one figure of merit is the squared norm of the displacement while the other is a regularization function. As to the latter, in} the open literature the $\ell_0$ norm of $\bm{I}$ has been employed, {and the bi-objective problem is handled} {by} minimizing it under {a} constraint {on the maximum amount of} displacement. In this regards, the CS framework {(with bespoke approximations of the $\ell_0$ norm)} has been successfully applied to this retrieval problem ({see} Subsection~\ref{sub:CS}).
An alternative {method} is to consider the rank of $\bm{I}$ as the regularization term and to {tackle} the optimization problem via {the} scalarization {framework~\cite{boyd2004convex}} ({see} Subsection~\ref{sub:minRank}).

\subsection{{Compressed-Sensing Recovery}}\label{sub:CS}
CS represents a powerful signal processing {tool} capable of reconstructing signals from a {small amount} of measurements~\cite{de2019compressed}. It has been successfully applied to accomplish ISAR imaging, yielding noteworthy improvements in the context of data reduction as well as resolution enhancement~\cite{tomei2016compressive}. The key principle underlying CS-based ISAR imaging lies in its ability to leverage the sparsity {assumption of} scatterers in the image domain. {From a physical point of view,} at high frequencies, {a radar signal is sparse in the image domain (i.e., when represented in the two-dimensional (2D) Fourier	basis), since it can be {accurately described} by the superimposition of few prominent scatterer responses in the radar image plane}~\cite{tomei2016compressive}.
	
{Accordingly,} the CS approach reconstructs the ISAR image $\bm{I}$ by solving the following optimization problem
	\begin{equation}\label{eq:problem_CS}
		\begin{aligned}
			\hat{\bm{I}} = &\operatorname{arg} \min\limits_{\bm{I}} \|\bm{I}\|_{0} \\
			& \text{ s.t. } \|\bm{S} - \bm{\Theta}_x \bm{I} \bm{\Theta}_y^\dagger \|_F^2 \le \epsilon,
		\end{aligned}
	\end{equation}
	where $\epsilon$ is a user-defined {tuning} parameter accounting for the noise level in the measured data.
	
	{Notably, in~\cite{tomei2016compressive}, the 2D-SL0 algorithm~\cite{ghaffari2009sparse} has been successfully employed to} {handle}~\eqref{eq:problem_CS}, obtaining {a sparse} solution consistent with the observed data. The procedure is {summarized in \textbf{Algorithm~\ref{alg:2DSL0}}, where, unless otherwise specified, $\sigma_{\min}=10^{-6}, \alpha=0.6, \mu_0 = 2$, and $L=15$.}

\begin{algorithm}[t]
	\caption{2D-SL0 Algorithm}\label{alg:2DSL0}
	\KwIn{$\bm{\Theta}_x$, $\bm{\Theta}_y$, $\bm{S}$, $\sigma_{\min}$, $\alpha$, $\mu_0$, $L$.}
	Compute $\bm{\Theta}_x^+$ and $ \bm{\Theta}_y^+$\;
	Compute the initial estimate: $\bm{I} = \bm{\Theta}_x^+ \bm{S} (\bm{\Theta}_y^+)^\dagger$\;
	Set $\sigma = 2 \, \|\bm{I}\|_{\textrm{max}}$\;
	\While{$\sigma > \sigma_{\min}$}{
		\For{$i = 1$ \KwTo $L$}{
			Update: $\bm{D} = \bm{I} \odot \bm{G}$, with $\bm{G}(m,n) = \exp\left(-\frac{|\bm{I}(m,n)|^2}{2\sigma^2}\right)$\;
			Gradient step: $\bm{I} = \bm{I} - \mu_0 \bm{D}$\;
			Projection: $\bm{E} = \bm{\Theta}_x^+ (\bm{\Theta}_x \bm{I} \bm{\Theta}_y^\dagger - \bm{S}) (\bm{\Theta}_y^+)^\dagger$\;
			Update: $\bm{I} = \bm{I} - \bm{E}$\;
		}
		Update $\sigma = \alpha \sigma$\;
	}
	
	\KwOut{$\hat{\bm{I}} = \bm{I}$}
\end{algorithm}

\subsection{Rank-Minimization (RM) Recovery}\label{sub:minRank}
{In this subsection, a different regularization function is considered, i.e., the rank of $\bm{I}$ which rules the amount of modes describing the image. As a consequence, the figure of merits involved in the inference problem are
\begin{itemize}
	\item $\| \bm{S} - \bm{\Theta}_x \bm{I} \bm{\Theta}_y^\dagger \|_F^2$, 
	which {controls} the fitting of the recovered image with the available measurements;
	\item $\rank\{\bm{I}\}$, 
	which accounts for the amount of modes required to {faithfully} describe the image.
\end{itemize}}

{Hence, to face with the resulting bi-objective optimization, as already mentioned, the scalarization framework~\cite{boyd2004convex} is exploited.}
Precisely, the ISAR {image retrieval} is formulated as
\begin{equation}\label{eq:problem_min_rank}
		\begin{aligned}
				\hat{\bm{I}} = &\operatorname{arg} \min\limits_{\bm{I}} 
				{\frac{1}{2}}\left\| \bm{S} - \bm{\Theta}_x \bm{I} \bm{\Theta}_y^\dagger \right\|_F^2 + \lambda \rank\{\bm{I}\},
			\end{aligned}
	\end{equation}
which is an unconstrained {single}-objective optimization problem with $\lambda$ the corresponding weight parameter that rules the trade-off between the displacement and the regularization terms.

It is important to observe that Problem~\eqref{eq:problem_min_rank} is NP-hard (due to the rank {term})~\cite{liu2014exact}. Hence, to determine a high quality sub-optimal solution, the rank operator is approximated via the plain nuclear norm, which is the closest convex relaxation {to the rank function}~\cite{5452187, fazel2002matrix, liu2014exact}. According to the above guidelines, the inference problem {becomes}
\begin{equation}\label{eq:problem_min_rank2}
	\begin{aligned}
		\hat{\bm{I}} = &\operatorname{arg} \min\limits_{\bm{I}} f(\bm{I}), \; 
		f(\bm{I}) \triangleq {\frac{1}{2}}\left\| \bm{S} - \bm{\Theta}_x \bm{I} \bm{\Theta}_y^\dagger \right\|_F^2 + \lambda \|\bm{I}\|_\star,
	\end{aligned}
\end{equation}
which is a convex optimization problem. A viable strategy {to minimize the objective in~\eqref{eq:problem_min_rank2}} relies on the MM framework, which is a powerful iterative procedure capable of addressing challenging optimization problems in an efficient and scalable way (see~\cite{sun2016majorization, razaviyayn2013unified, aubry2018new} for technical details). 
The method involves the construction of a tailored surrogate function  (s.f.) that satisfies specific properties, leveraging {on} the optimized solution from the previous iteration. The s.f. is then minimized to update the optimized point.

{To proceed further, let us denote} by $\bm{I}_k$ the estimate of $\bm{I}$ at the $k$th iteration{;} following {similar} line of reasoning as in~\cite{liu2014exact}, a {s.f.} $g(\bm{I};\bm{I}_k)$ for the objective function in~\eqref{eq:problem_min_rank2} is given by the quadratic approximation of $f(\bm{I})$, i.e.,
\begin{equation}\label{eq:s.f.}
	g(\bm{I};\bm{I}_k) = \frac{\tau}{2} \|\bm{I}-\bar{\bm{I}}_k\|_F^2 + \lambda \|\bm{I}\|_\star + f(\bm{I}_k) - \frac{1}{2\tau} \|\nabla f(\bm{I}_k)\|_F^2
\end{equation}
{where $\tau \ge \mathcal{L}_f \triangleq \lambda_{max}(\bm{C}^\dagger \bm{C}) = \lambda_{max}(\bm{\Theta}_x^\dagger \bm{\Theta}_x) \lambda_{max}(\bm{\Theta}_y^\dagger \bm{\Theta}_y)$~\cite{liu2014exact}, with $\bm{C} = \bm{\Theta}_y^\star \otimes \bm{\Theta}_x$,}
\begin{equation}
	\bar{\bm{I}}_k = \bm{I}_k - \frac{1}{\tau} \bm{\Theta}_x^\dagger \left(\bm{\Theta}_x \bm{I}_k \bm{\Theta}_y^\dagger - \bm{S} \right)\bm{\Theta}_y,
\end{equation}
whereas $\nabla f(\bm{I}_k)$ is the gradient of $f(\bm{I})$ evaluated at $\bm{I}_k$.

{Let us now focus on the minimization of} $g(\bm{I};\bm{I}_k)$. To this end, let $\bm{U}_k \bm{\Sigma}_k \bm{V}_k^\dagger$ be the Singular Value Decomposition (SVD) of $\bar{\bm{I}}_k$ and $ \bm{U}\bm{\Sigma}\bm{V}^\dagger$ that of $\bm{I}$, with the singular values in $\bm{\Sigma}_k$ and $\bm{\Sigma}$ sort in descending order{; hence}, based on~\cite[Corollary 7.4.1.5]{horn2012matrix},
	\begin{equation}
		\| {\bm{I}}-\bar{\bm{I}}_k \|_F^2 \ge \|{\bm{\Sigma}} - \bm{\Sigma}_k\|_F^2
	\end{equation}
	with the equality if $\bm{I} = \bm{U}_k \bm{\Sigma} \bm{V}_k^\dagger$.
As a consequence, the left and right singular vectors of {a minimizer {for} the objective in}~\eqref{eq:s.f.} are the same {as} $\bar{\bm{I}}_k$. Therefore, assuming that $\bar{\bm{I}}_k$ {has} $r$ non-zero singular values  $\{\bar{\sigma}_k^{(i)}\}_{i=1}^r$,  the optimization problem at hand {reduces} to computing the singular values {as solution of the optimization problem}
\begin{equation}\label{eq:sf2}
{\min_{\sigma^{(1)}\ge 0, \dots, \sigma^{(r)}\ge 0}}	\tilde{g}(\bm{I};\bm{I}_k) \triangleq \sum_{i=1}^{r} \left[\frac{\tau}{2}\left( \sigma^{(i)} - \bar{\sigma}_k^{(i)} \right)^2 + \lambda \sigma^{(i)}\right]
\end{equation}
{where $\sigma^{(i)}, i=1,\dots,r$, are the possible non zero singular values of $\bm{I}$.} {Additionally}, since the problem is separable in its variables, the $i$th singular value of $\bm{I}$ can be computed solving
\begin{equation}\label{eq:problem_sigma}
	{\min_{\sigma^{(i)}\ge 0} \frac{\tau}{2}\left( \sigma^{(i)} - \bar{\sigma}_k^{(i)} \right)^2 + \lambda \sigma^{(i)}}, \; i=1, \dots, r,
\end{equation}
whose optimal value is given by
\begin{equation}
	\sigma^{(i)} = \left(\bar{\sigma}_k^{(i)} - \frac{\lambda}{\tau}\right)_+ .
\end{equation}

{Summarizing}, a minimizer of~\eqref{eq:s.f.} is
\begin{equation}
	{\bm{I}}_{k+1} = \bm{U}_k \bm{{\Sigma}}_{k+1} \bm{V}_k^\dagger
\end{equation}
with
\begin{equation}
	\bm{{\Sigma}}_{k+1} = \left(\bm{\Sigma}_{k} - \frac{\lambda}{\tau} \bm{I}\right)_+ .
\end{equation}

The {overall} procedure is detailed in \textbf{Algorithm~\ref{alg:MM}}, {wherein $\bm{I}_0$ is the initialization matrix (e.g., the output of the 2D-SL0 algorithm)}. {Finally}, {denoting by $\bm{I}_M = \bm{\Theta}_x^+ \bm{S} \left(\bm{\Theta}_y^{\dagger}\right)^+ $ the ISAR image computed on the incomplete data,} the procedure terminates when either the condition $\operatorname{rank}_S\{\bm{I}_k\} \le \operatorname{rank}_S\{\bm{I}_M\}$ is met or the maximum number of iterations $k_{\operatorname{MAX}}$ is reached, {where $\srank\{\bm{I}\} = \exp \left[-\sum_{i=1}^r \frac{\sigma_i(\bm{I})}{\|\bm{\sigma}\|_1} \log \left(\frac{\sigma_i(\bm{I})}{\|\bm{\sigma}\|_1}\right)\right]$ is the Shannon effective rank of $\bm{I}$~\cite{roy2007effective}, with $\bm{\sigma} = [\sigma_1(\bm{I}), \dots, \sigma_r(\bm{I})]^\mathrm{T}$ the vector comprising the singular values of $\bm{I}$.}

\begin{algorithm}[t]
	\caption{MM-based RM algorithm}\label{alg:MM}
	\KwIn{$\bm{I}_0$, $\lambda$, $\bm{S}$, $\bm{\Theta}_x$, $\bm{\Theta}_y$, $k = k_{\operatorname{MAX}}$.}

	set $\tau \ge \lambda_{max}(\bm{\Theta}_x^\dagger \bm{\Theta}_x) \lambda_{max}(\bm{\Theta}_y^\dagger \bm{\Theta}_y)$\;
	set $k=0$\;
	set {$\bm{I}_M = \bm{\Theta}_x^+ \bm{S} \left(\bm{\Theta}_y^{\dagger}\right)^+ $}\;
	\Repeat{{$\operatorname{rank}_S\{\bm{I}_k\} \le \operatorname{rank}_S\{{\bm{I}_M}\}$ or $k = k_{\operatorname{MAX}}$}}{
		set $\bar{\bm{I}}_k = \bm{I}_k - \frac{1}{\tau} \bm{\Theta}_x^\dagger \left(\bm{\Theta}_x \bm{I}_k \bm{\Theta}_y^\dagger - \bm{S} \right)\bm{\Theta}_y$\;
		compute the SVD of $\bar{\bm{I}}_k$: $\bar{\bm{I}}_k = \bm{U}_k \bm{\Sigma}_k \bm{V}_k^\dagger$\;
		set $\bm{\Sigma}_{k+1} = \left(\bm{\Sigma}_{k} - \frac{\lambda}{\tau} \bm{I}\right)_+$\;
		set $\bm{I}_{k+1} = \bm{U}_k \bm{\Sigma}_{k+1} \bm{V}_k^\dagger$\;
		set $k = k+1$\;
	} 
	\KwOut{$\hat{\bm{I}} = \bm{I}_k$}
\end{algorithm}

\subsection{RD-based Imaging}
After the reconstruction stage performed by means of the 2D-SL0 or {\bf Algorithm~\ref{alg:MM}}, leveraging the conventional Fourier dictionaries $\bm{\Psi}_x$ and $\bm{\Psi}_y$ that {account for} the range compression and the cross-range compression, respectively, it is possible to reconstruct the slow-time/frequency domain data {matrix} as
\begin{equation}
	\hat{\bm{S}}_c = \bm{\Psi}_x \hat{\bm{I}} \bm{\Psi}_y^\dagger
\end{equation}
which allows to {realize} the ISAR {imaging} by means of the conventional RD algorithm applied on $	\hat{\bm{S}}_c$~\cite{tomei2016compressive}.
	
	\section{Numerical Analysis}

	\begin{figure}[t]
		\centering
		\includegraphics[width=0.99\linewidth]{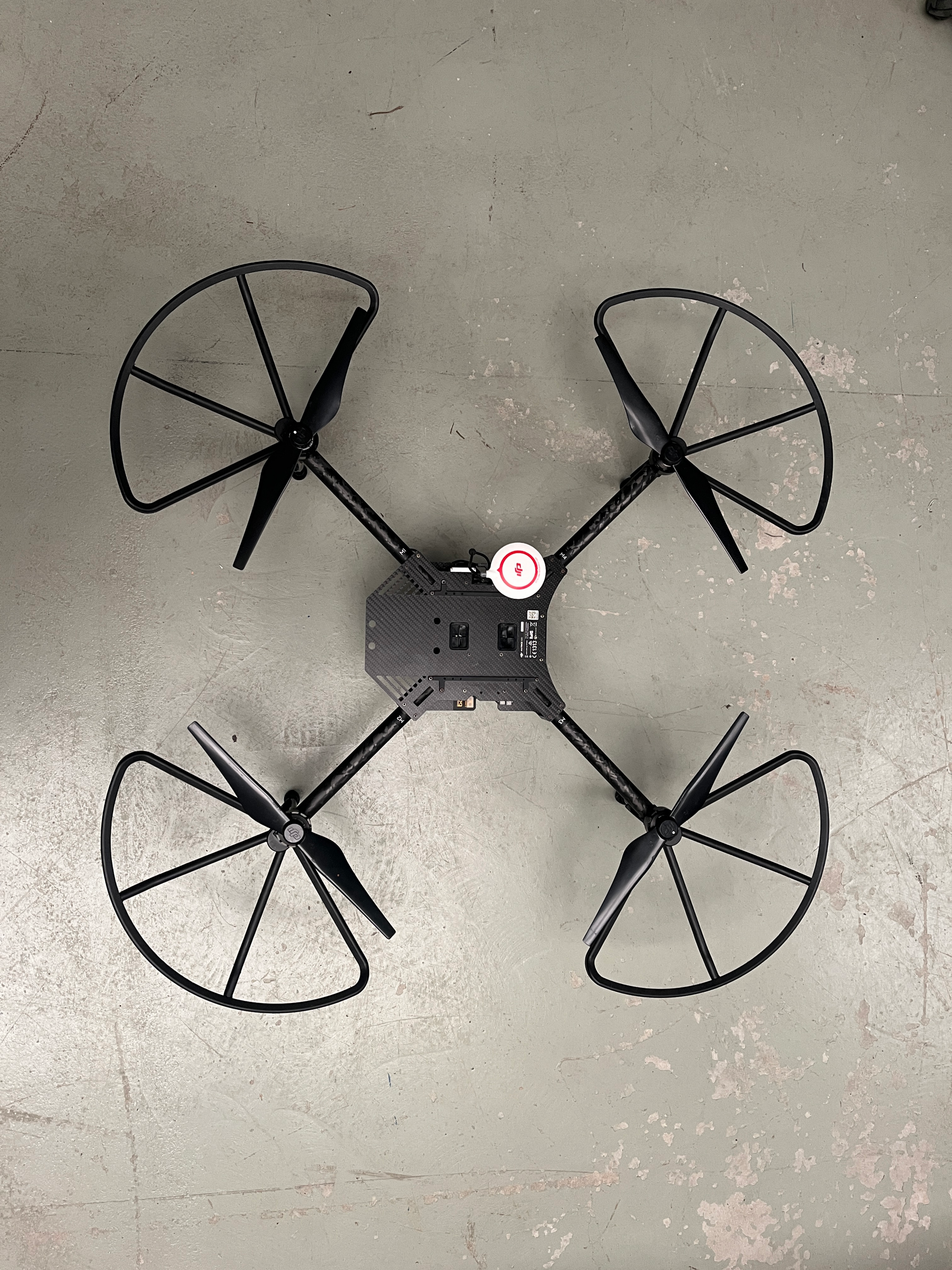}
		\caption{DJI Matrice 100 {used to collect} measure data~\cite{9982651}.}
		\label{fig:djimatrice100}
	\end{figure}
		
	\begin{figure*}[t]
		\centering
		\includegraphics[width=0.99\linewidth]{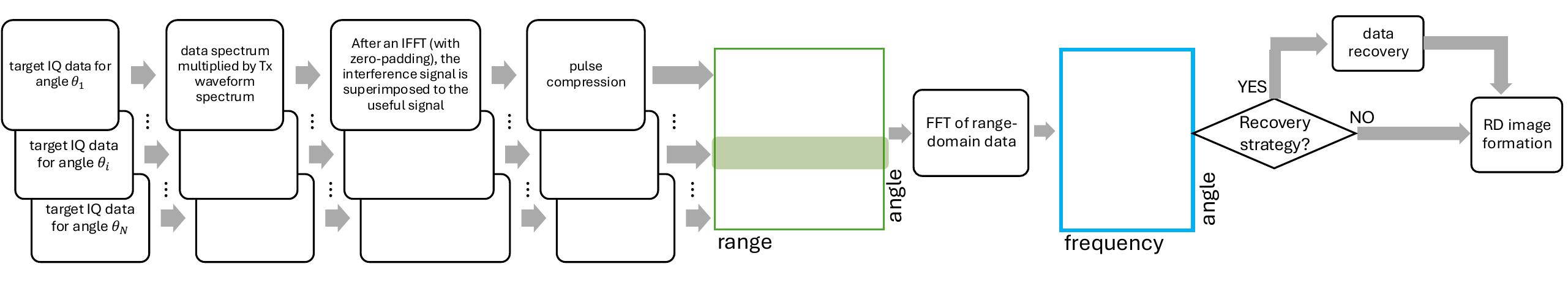}
		\caption{{Flowchart depicting the {scenario generation} process employed for the numerical analysis of Section IV}.}
		\label{fig:flowchart}
	\end{figure*}

	In this section, the effectiveness of the proposed {procedure} is investigated by comparing the image obtained with the proposed cognitive ISAR process with those resulting from an ideal case {(receiving only the useful target contribution {without spectral holes}) as well as a standard ISAR operating in the presence of interference}. {The} comparison is conducted both visually and numerically by resorting to specific {quantitative} metrics such as 
	\begin{itemize}
		\item {Image Contrast (IC)}:  
		It evaluates the relative intensity variations in the image, providing an indication of its overall clarity and sharpness. It is defined as~\cite{532282, 9218965}
		\begin{equation}
			IC(\bm{I}) = \frac{\sqrt{\mathbb{A}[(|\bm{I}| - \mathbb{A}[|\bm{I}|])^2]}}{\mathbb{A}[|\bm{I}|]}.
		\end{equation}
		
		\item {Image Coherence (COH)}:  
		It computes the similarity between two images, i.e., the resulting ISAR image $\bm{I}$ obtained with the proposed method and a reference image $\bm{I}_R$ (ideal case without missing-data). It is given by:
		\begin{equation}
			COH(\bm{I}, \bm{I}_R) = \frac{\left|\bm{1}^\mathrm{T} \left(\bm{I}_R \odot \bm{I}^\star\right) \bm{1}\right|}{\|\bm{I}\|_F \; \|\bm{I}_R\|_F}.
		\end{equation}
		
		\item {Normalized Mean Squared Error (NMSE)}:  
		This metric represents the normalized error between the obtained image and the reference
		\begin{equation}
			NMSE(\bm{I}, \bm{I}_R) = \frac{\|\bm{I}- \bm{I}_R\|_F}{\| \bm{I}_R \|_F}.
		\end{equation}
	\end{itemize}

	\begin{figure*}[h!]
		\centering  
		\begin{minipage}[b]{0.241\linewidth} 
			\centering   
			\includegraphics[width=\linewidth]{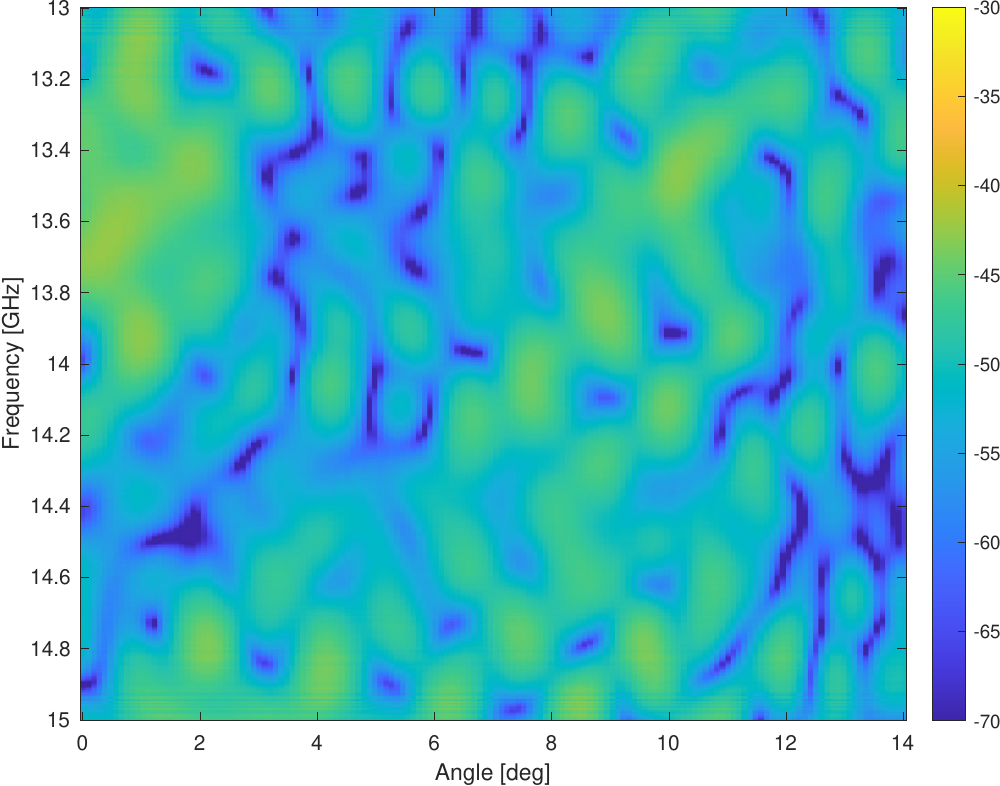}
			\subcaption{}\end{minipage}\hfill 
		\begin{minipage}[b]{0.241\linewidth}   
			\centering   
			\includegraphics[width=\linewidth]{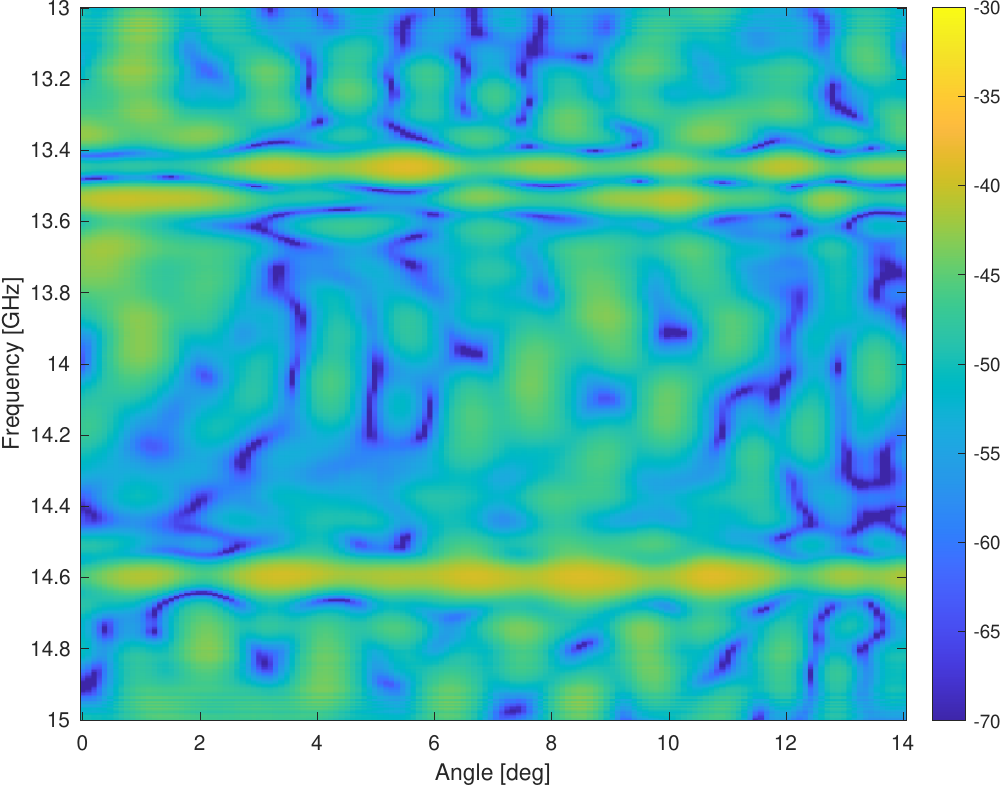}   
			\subcaption{}\end{minipage}\hfill 
		\begin{minipage}[b]{0.241\linewidth}   
			\centering   
			\includegraphics[width=\linewidth]{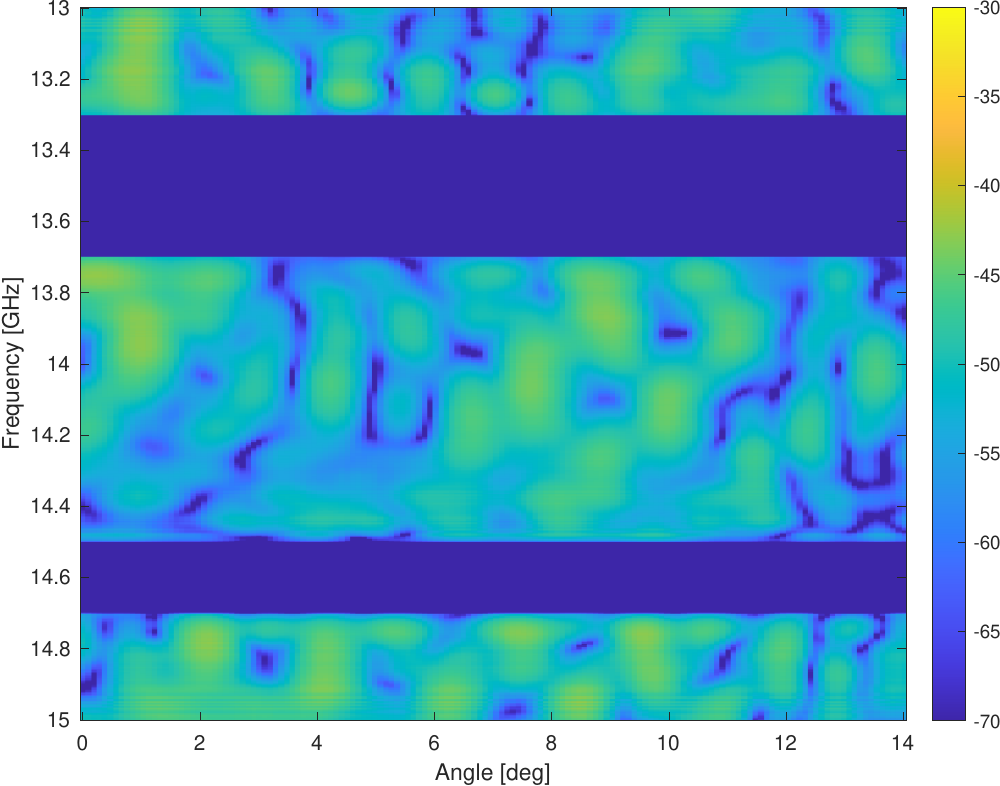}   
			\subcaption{}\end{minipage}\hfill 
		\begin{minipage}[b]{0.241\linewidth}   
			\centering   
			\includegraphics[trim=0 0 0 10,clip,width=\linewidth]{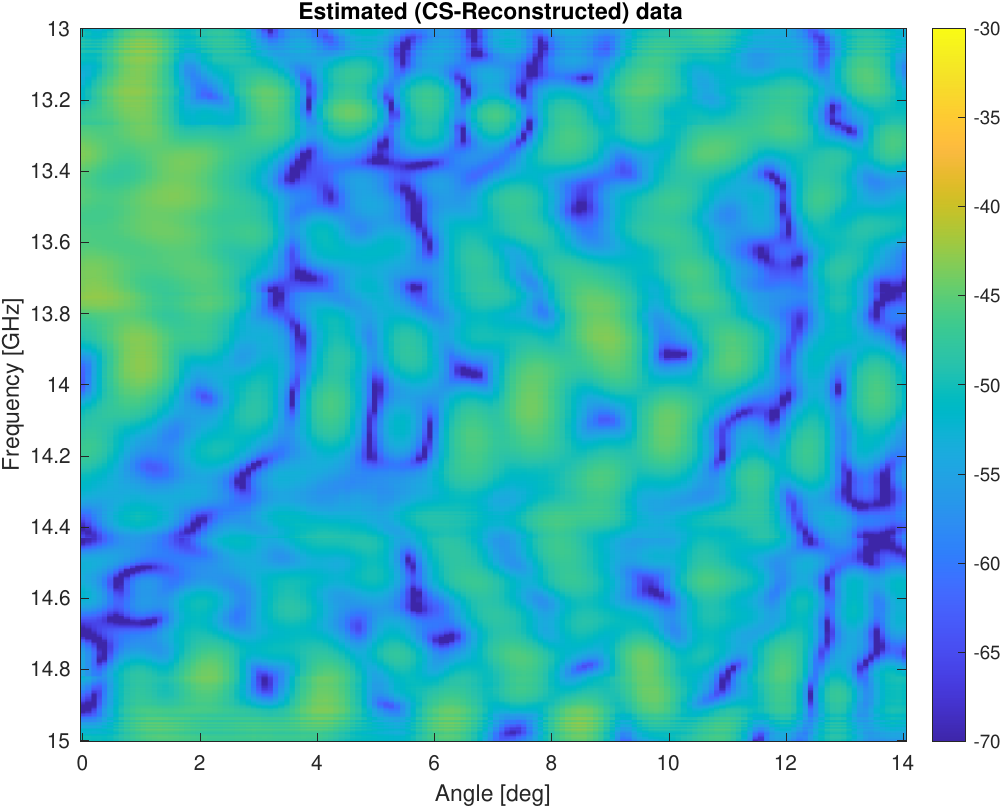}   
			
			\subcaption{}\end{minipage}\hfill \caption{Frequency-angle (dB) spectrum in the presence of two interferers. (a) GT (b) {standard case} (c) {notched case} (d) {N-CS/N-RM} case.}
		\label{fig:map}
	\end{figure*}
	\begin{figure*}[h!]
		\centering  
		\begin{minipage}[b]{0.241\linewidth}   
			\centering   
			\includegraphics[width=\linewidth]{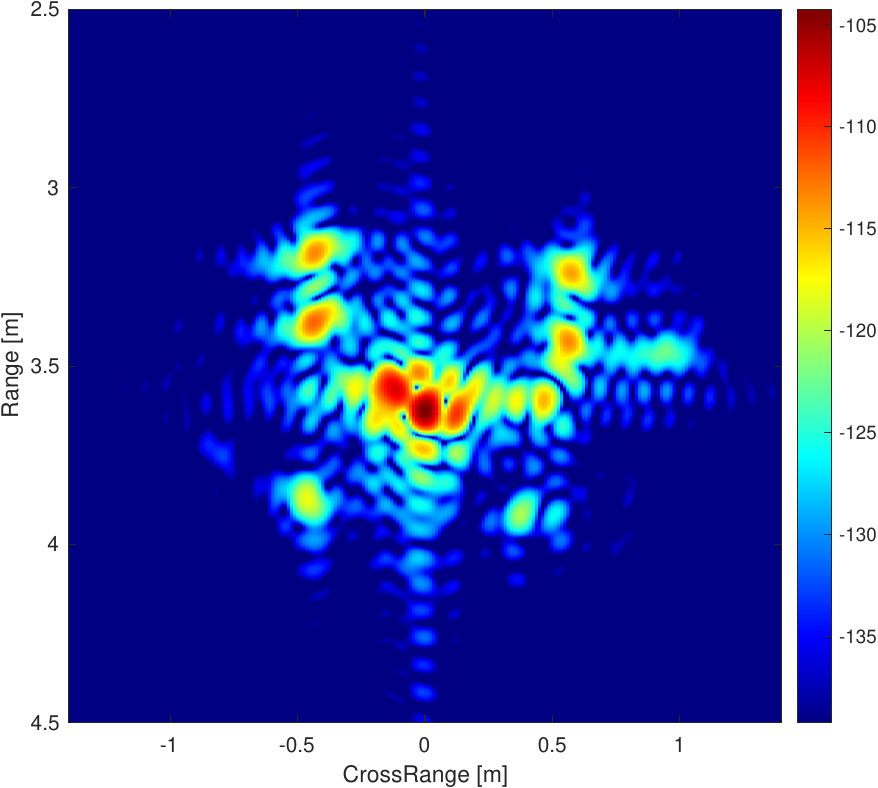}   
			
			\subcaption{}\end{minipage}\hfill 
		\begin{minipage}[b]{0.241\linewidth}   
			\centering   
			\includegraphics[width=\linewidth]{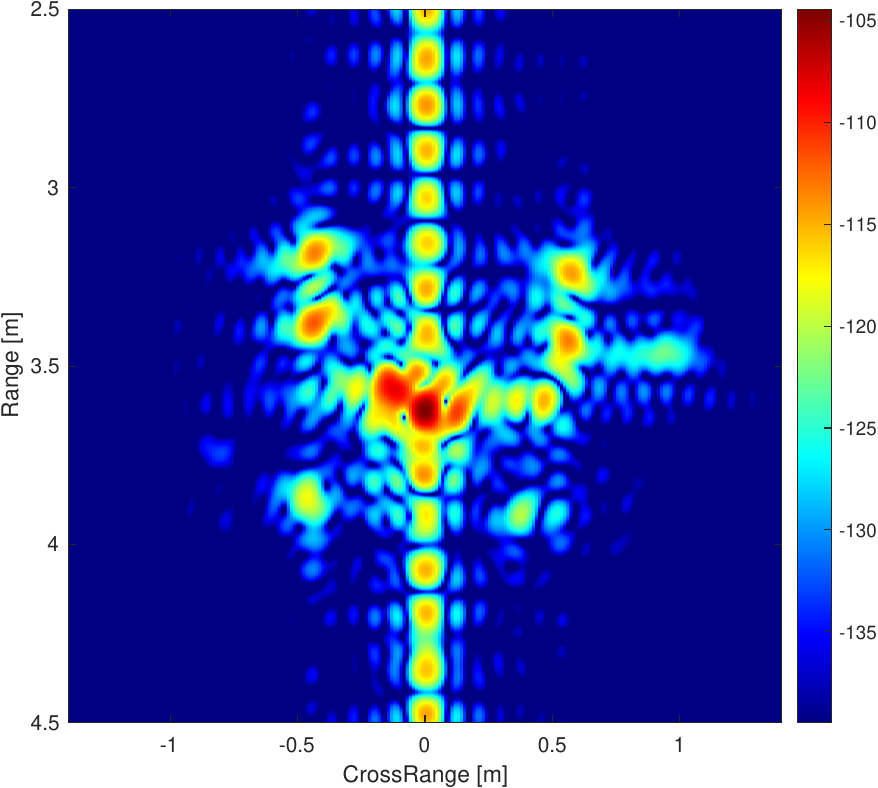}   
			
			\subcaption{}\end{minipage}\hfill 
		\begin{minipage}[b]{0.241\linewidth}   
			\centering   
			\includegraphics[width=\linewidth]{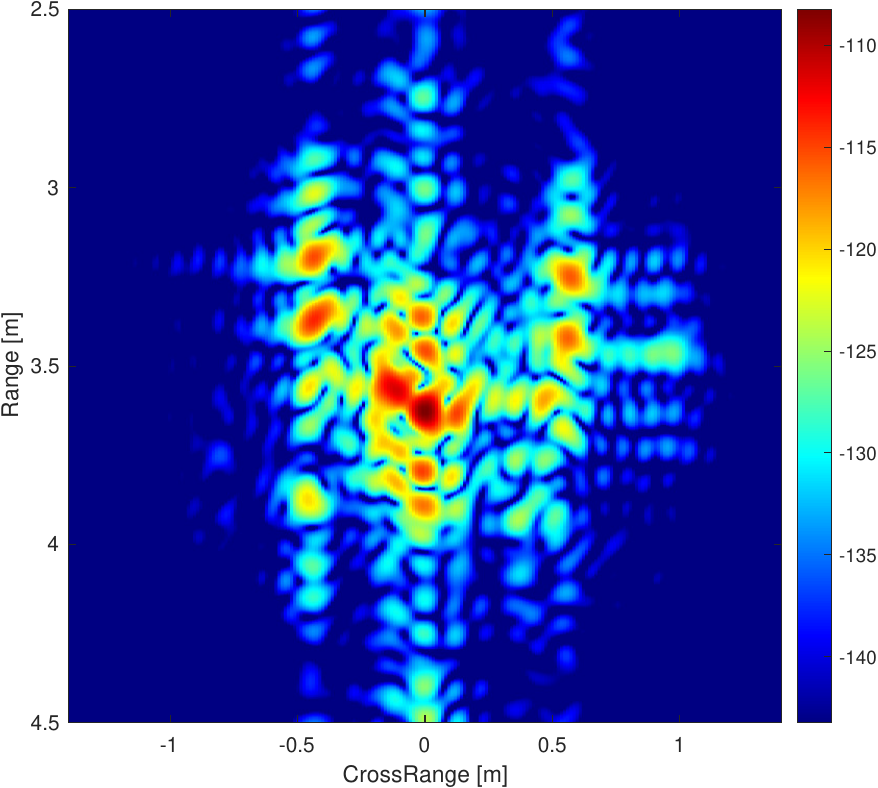}   
			
			\subcaption{}\end{minipage}\hfill 
		\begin{minipage}[b]{0.241\linewidth}   
			\centering   
			\includegraphics[width=\linewidth]{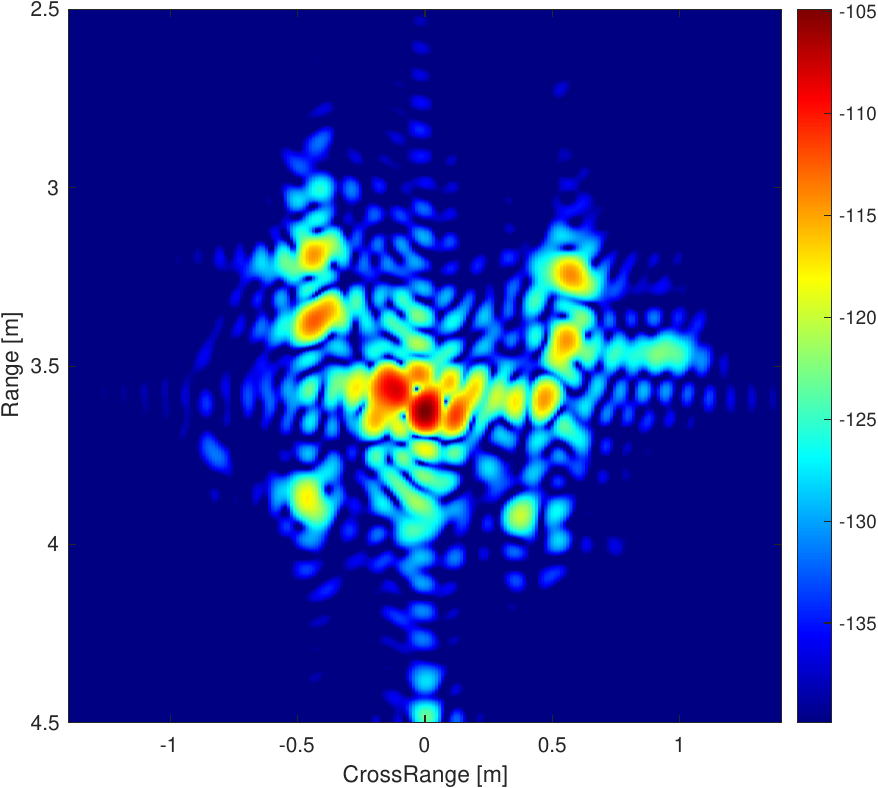}   
			
			\subcaption{}\end{minipage}\hfill \caption{{ISAR image (dB)} in the presence of two interferers. (a) GT (b) {standard case} (c) {notched case} (d) {N-CS/N-RM} case.}
		\label{fig:ISAR}
	\end{figure*}
	
	The dataset utilized for the analysis is the one detailed in~\cite{9982651}, which comprises measurements of various drones acquired in a semi-controlled environment across a frequency range of 8.2-18 GHz, using HH and VV polarizations, and azimuth aspect angles spanning 0-360 degrees with 0.1 degree steps and elevation close to 0 degree. In the following, the data pertaining to the DJI Matrice 100 (illustrated in Fig.~\ref{fig:djimatrice100}), obtained in HH polarization with a 2 GHz bandwidth from 13 GHz to 15 GHz {(considering a frequency step of 4.5 MHz)} and with a target rotation of 15 degrees {(with an angular resolution of 0.1 degrees)}, {are} employed for the analysis.

	Five different {situations} are considered and compared:
	\begin{itemize}
		\item the ideal case, used as a ground truth (GT), which assumes the transmission of a standard linear frequency modulated (LFM) signal (chirp) in the absence of interference;
		\item the {standard case}, involving the transmission of {a} standard chirp in the presence of emitters operating within the same band as the ISAR frequency range;
		\item {the} notched case, comprising a perception stage, where the radar recognizes the presence of overlaid emitters, and an action stage, involving design and transmission of a tailored chirp-like signal with spectral notches {in the frequency spectra};
		\item {notched} + CS case (referred to as {N-CS} case), wherein the 2D-SL0 algorithm~\cite{ghaffari2009sparse} is applied to the data obtained in the {notched case} to reconstruct the frequency-angle data, thus allowing to counter the potential resolution loss induced by the waveform spectral notches.
		\item {{notched} + RM case (referred to as {N-RM} case), where \textbf{Algorithm~\ref{alg:MM}} is employed as recovery method, considering $\lambda = 300$, $\tau = 20 \lambda_{max}(\bm{\Theta}_x^\dagger \bm{\Theta}_x) \lambda_{max}(\bm{\Theta}_y^\dagger \bm{\Theta}_y)$, and $k_{\operatorname{MAX}} = 50$.}
	\end{itemize}
	In the aforementioned scenarios, all the overlaid sources are assumed to be {telecommunication} emitters {(representing the only disturbance contributions)}.

	For the aim of this analysis, the spectrum of each complex high-range-resolution (HRR) target echo signal (provided by the dataset~\cite{9982651} {using the same pre-processing steps}), for a given aspect angle $\theta_i$ (corresponding to a specific slow-time snapshot), is first multiplied by the transmitted waveform spectrum, {as {initial} step to model (approximately) the {noise-free} data collected by the ISAR receiver}. Subsequently, the range-domain signal, obtained following an inverse Fourier transform of the resulting spectrum, is superimposed on the interference. The resulting samples, which represent the signal measured by an actual ISAR system, undergo pulse compression with the transmitted waveform and are stored in the $i$th row of a range-angle data matrix. Once all the slow-time snapshots have been processed, {after computing the FFT of the range-domain data, the resulting frequency-angle} matrix is employed for the formation of the ISAR image using the standard RD algorithm. It should be noted that in the CS and RM cases, prior to the image formation step, {the data matrix undergoes a {frequency-angle} data recovery procedure, according to the methods detailed in Section~\ref{sub:CS} and \ref{sub:minRank}, respectively}. Finally, the obtained data {experience} the RD image formation. The sequence of the employed steps is illustrated in the flowchart reported in Fig.~\ref{fig:flowchart}.

	{For the synthesis of the spectrally-shaped waveform to transmit in the {notched case}}, {\textbf{Algorithm~\ref{alg:qcqp}} is used}, considering notch depths of 30 dB for the first notch and 40 dB for the second one, respectively, {with $\bar{N}=5000$ and $W=2500$.}

	The five mentioned cases are analyzed and compared across three operational scenarios, reporting the {amplitude} (dB) of the frequency-time (i.e., frequency-angle) data, the corresponding ISAR image (obtained using the standard RD imaging procedure), and the {aforementioned figures of merit} {to assess} ISAR image quality.
	
	The first scenario (Figs.~\ref{fig:map} and \ref{fig:ISAR}) considers two communication sources, with the former operating in the range $[13.38, 13.62]$ GHz with a $240$ MHz bandwidth, and the latter in the range $[14.53, 14.65]$ GHz with a $120$ MHz bandwidth. 
	
	The figures clearly show that, if not properly {accounted for}, interfering signals introduce noticeable spurious in the resulting ISAR image (Fig.~\ref{fig:ISAR} (b)). Employing the cognitive approach {only} and down-stream the pulse compression, the frequency-angle data results being gapped in the frequency domain (Fig.~\ref{fig:map} (c)). The presence of {these} missing data is {highly damaging} for the imaging process, resulting in high sidelobes {appearing as} fake scatterers (Fig.~\ref{fig:ISAR} (c)). Conversely, leveraging the {CS/RM} technique {(the two recovery strategies achieve indistinguishable results)}, it is possible to reconstruct the frequency-angle data (Fig.~\ref{fig:map} (d)) and obtain an high-quality ISAR image (Fig.~\ref{fig:ISAR} (d)), with negligible {visual} differences from the GT (Fig.~\ref{fig:ISAR} (a)).

	\begin{figure*}[tbp]
		\centering  
		\begin{minipage}[b]{0.241\linewidth}   
			\centering   
			\includegraphics[width=\linewidth]{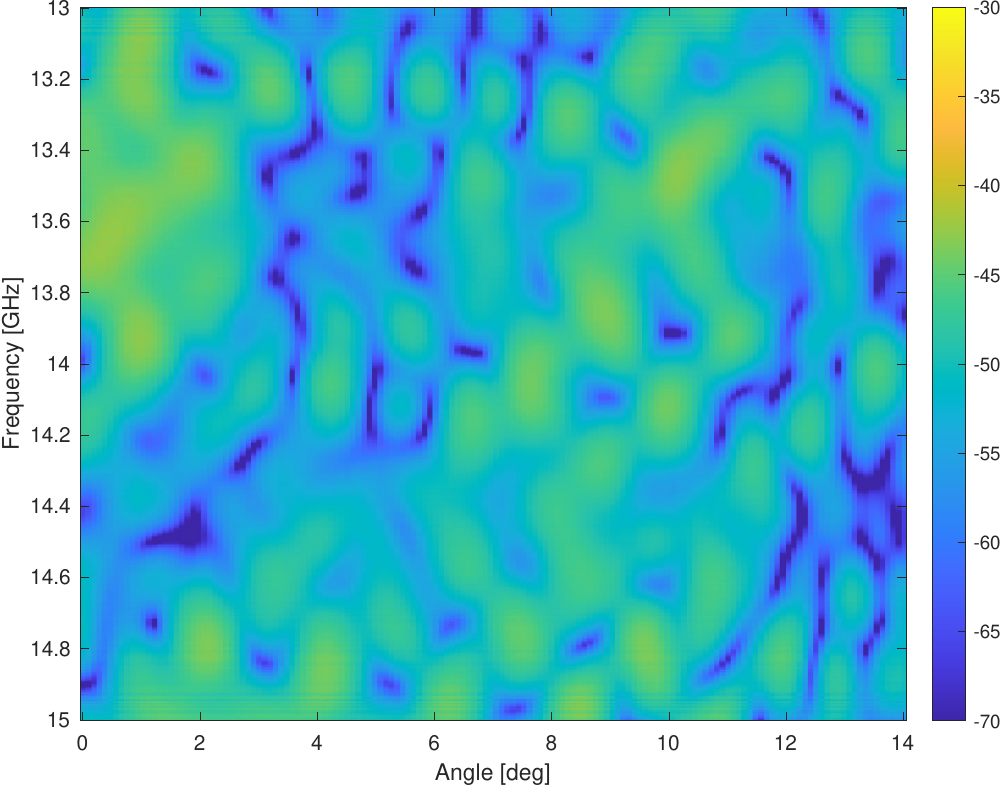}   
			
			\subcaption{}\end{minipage}\hfill 
		\begin{minipage}[b]{0.241\linewidth}   
			\centering   
			\includegraphics[width=\linewidth]{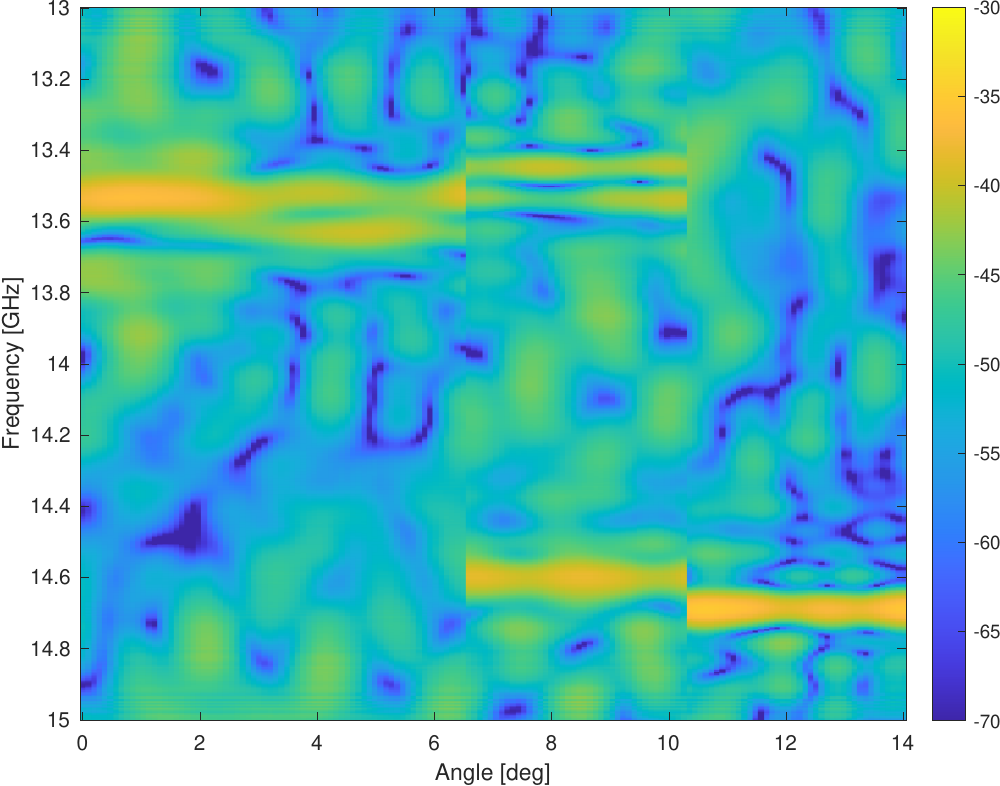}   
			
			\subcaption{}\end{minipage}\hfill 
		\begin{minipage}[b]{0.241\linewidth}   
			\centering   
			\includegraphics[width=\linewidth]{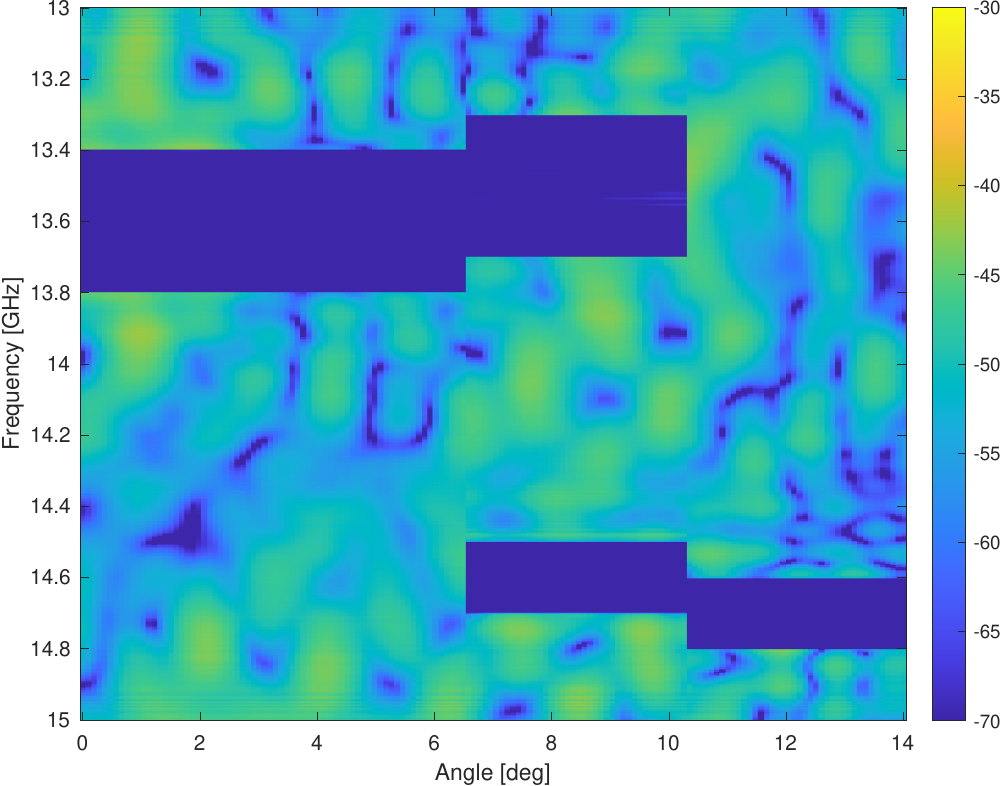}   
			
			\subcaption{}\end{minipage}\hfill 
		\begin{minipage}[b]{0.241\linewidth}   
			\centering   
			\includegraphics[trim=0 0 0 10,clip,width=\linewidth]{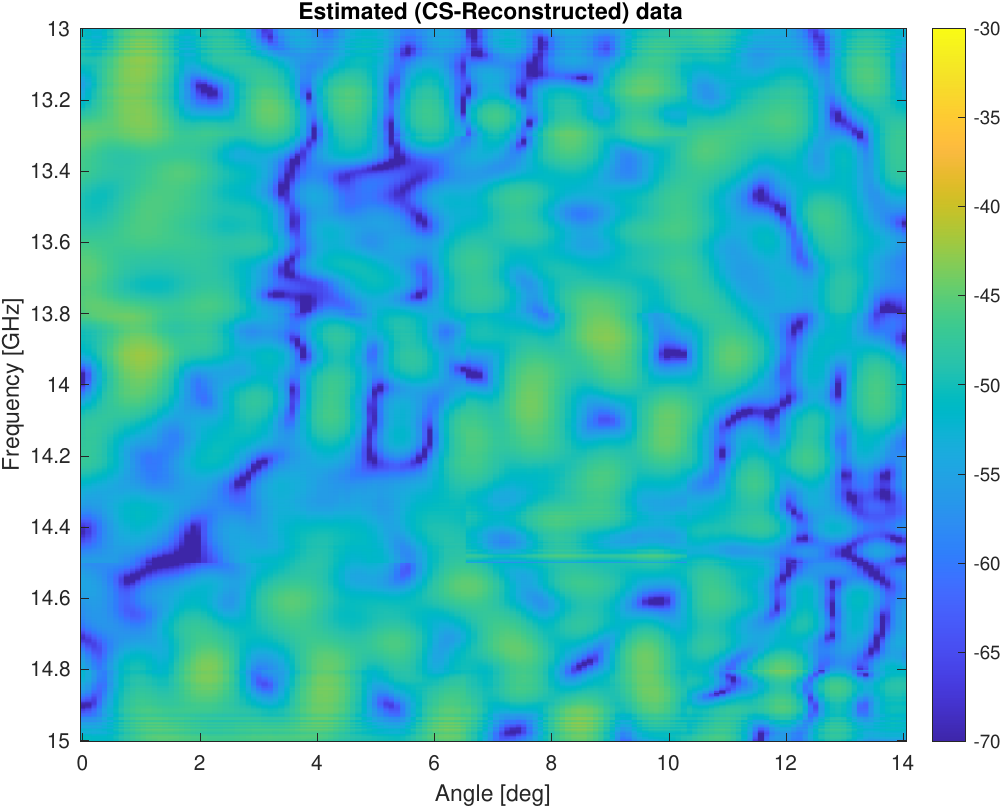}   
			
			\subcaption{}\end{minipage}\hfill \caption{Frequency-angle (dB) spectrum in the presence of {multiple emitters active in different temporal slots}. (a) GT (b) {standard case} (c) {notched case} (d) {N-CS/N-RM} case.}
		\label{fig:map_ag}
	\end{figure*}
	\begin{figure*}[h!]
		\centering  
		\begin{minipage}[b]{0.241\linewidth}   
			\centering   
			\includegraphics[width=\linewidth]{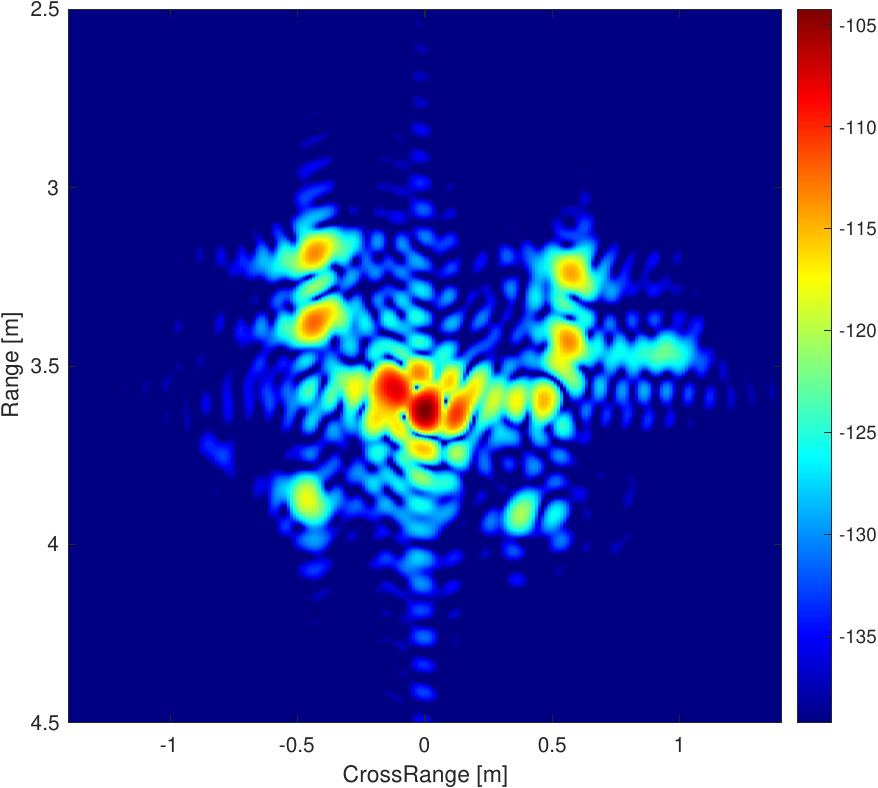}   
			
			\subcaption{}\end{minipage}\hfill 
		\begin{minipage}[b]{0.241\linewidth}   
			\centering   
			\includegraphics[width=\linewidth]{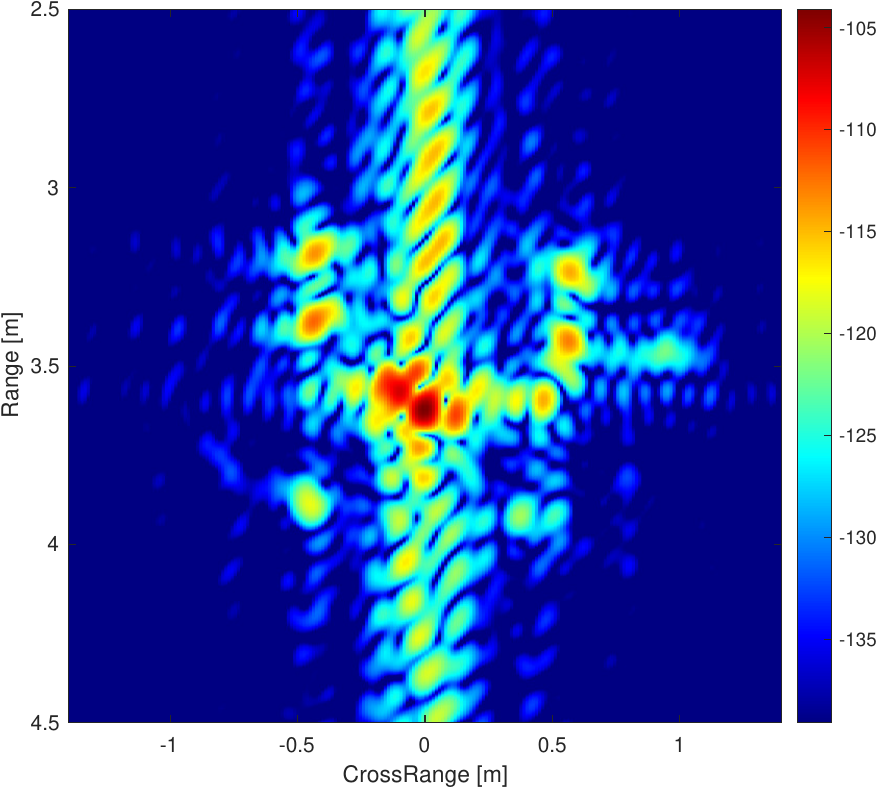}   
			
			\subcaption{}\end{minipage}\hfill 
		\begin{minipage}[b]{0.241\linewidth}   
			\centering   
			\includegraphics[width=\linewidth]{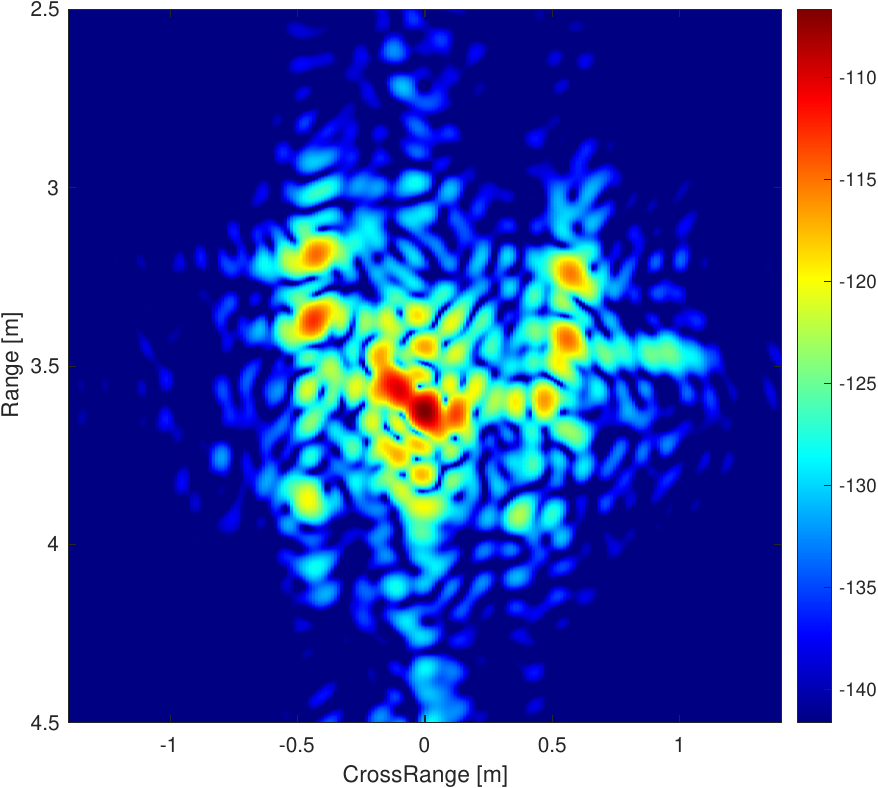}   
			
			\subcaption{}\end{minipage}\hfill 
		\begin{minipage}[b]{0.241\linewidth}   
			\centering   
			\includegraphics[width=\linewidth]{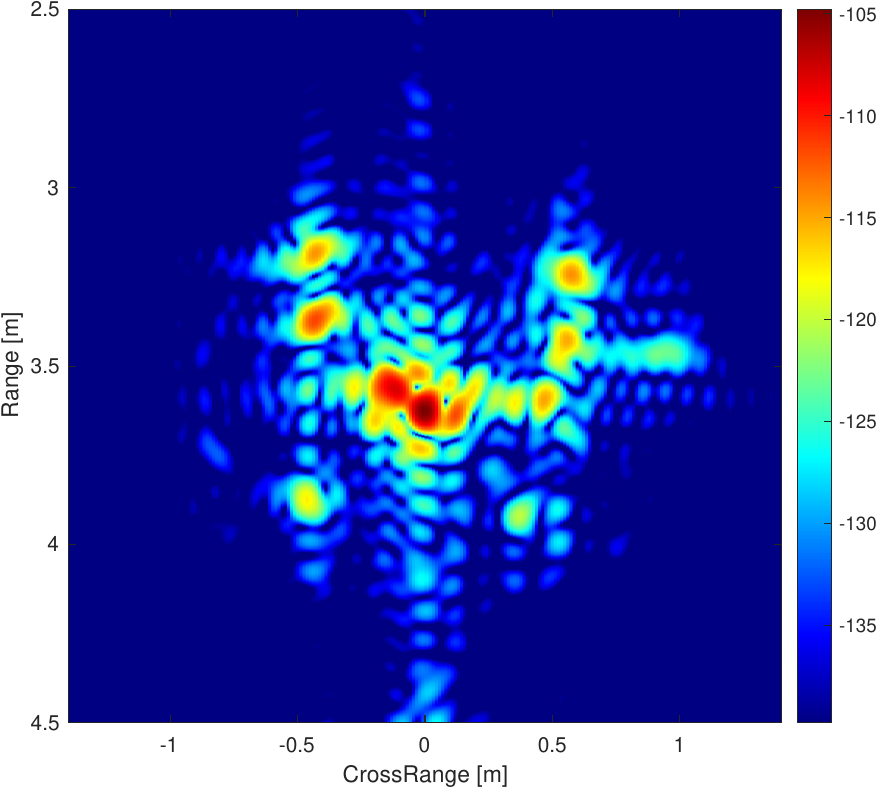}   
			
			\subcaption{}\end{minipage}\hfill \caption{{ISAR image (dB)} in the presence of {multiple emitters active in different temporal slots}. (a) GT (b) {standard case} (c) {notched case} (d) {N-CS/N-RM} case.}
		\label{fig:ISAR_ag}
	\end{figure*}
	
	{The second scenario (illustrated in Figs.~\ref{fig:map_ag} and \ref{fig:ISAR_ag}) considers the presence of multiple emitters, each active only during specific temporal slots corresponding to distinct target aspect angle intervals:
	\begin{itemize}
		\item 	A single emitter operates in the $[13.45, 13.69]$ GHz band (240 MHz bandwidth) during the interval associated with target aspect angles in the range $[0^\circ, 7^\circ]$.
		\item 	Two emitters are active in the subsequent interval, covering aspect angles from $[7^\circ, 11^\circ]$: the former transmits in the $[13.38, 13.62]$ GHz band (240 MHz bandwidth), and the latter in the $[14.53, 14.65]$ GHz band (120 MHz bandwidth).
		\item 	Finally, a single emitter transmits in the $[14.665, 14.745]$ GHz band (80 MHz bandwidth) during the interval corresponding to aspect angles in the range $[11^\circ, 15^\circ]$.
	\end{itemize}
}	
			
	
	{Like in} the previous scenario, the plots reveal that only in the {{N-CS}/{N-RM}} case it is possible to obtain a {faithful} ISAR image of the drone. As a matter of fact, in the other cases, the interference and the missing data  noticeable {degrade the image quality}, hindering the recognition of the drone shape.

\begin{figure*}[h!]
		\centering  
		\begin{minipage}[b]{0.241\linewidth}   
			\centering   
			\includegraphics[width=\linewidth]{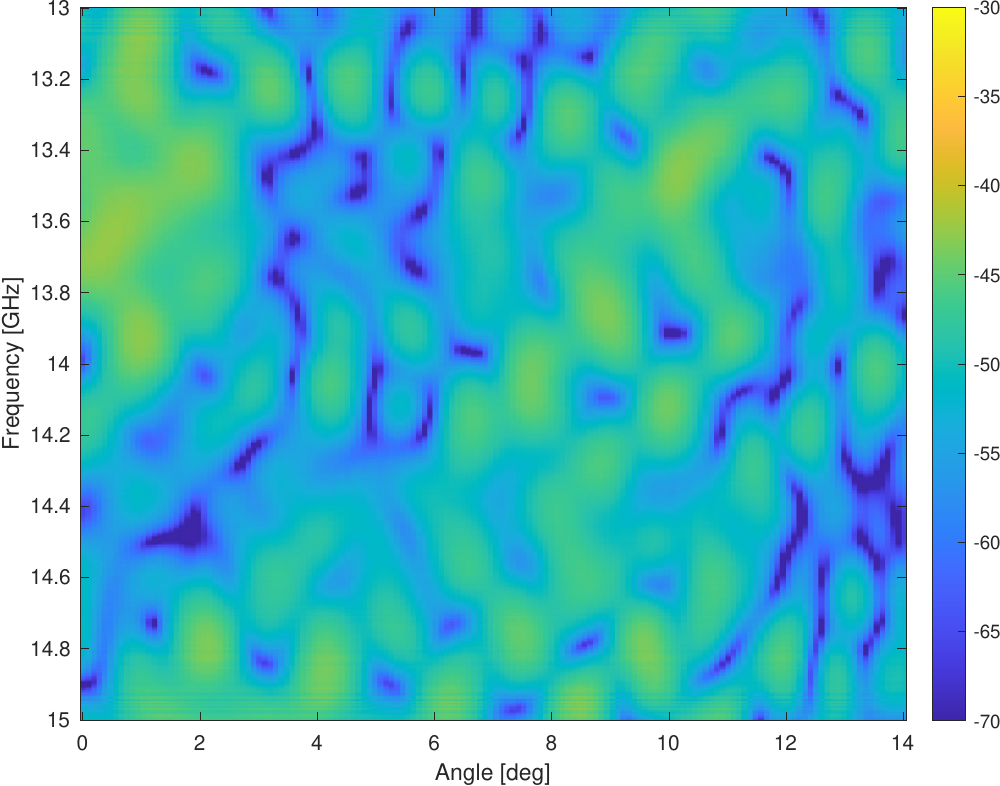}   
			
			\subcaption{}\end{minipage}\hfill 
		\begin{minipage}[b]{0.241\linewidth}   
			\centering   
			\includegraphics[width=\linewidth]{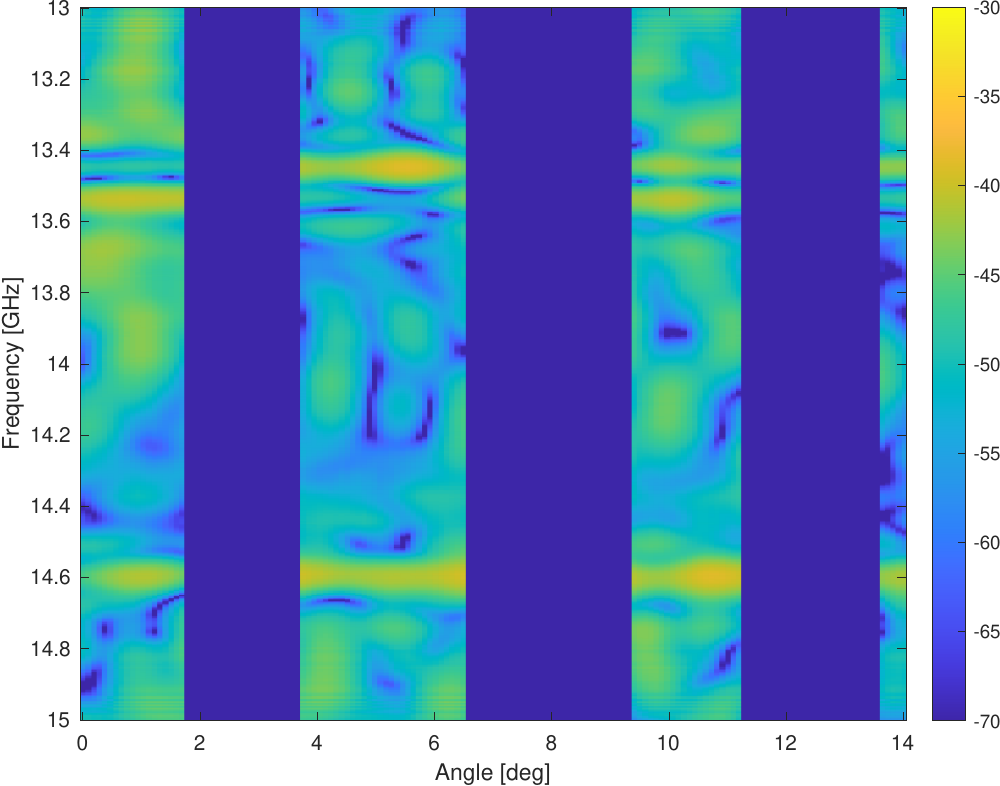}   
			
			\subcaption{}\end{minipage}\hfill 
		\begin{minipage}[b]{0.241\linewidth}   
			\centering   
			\includegraphics[width=\linewidth]{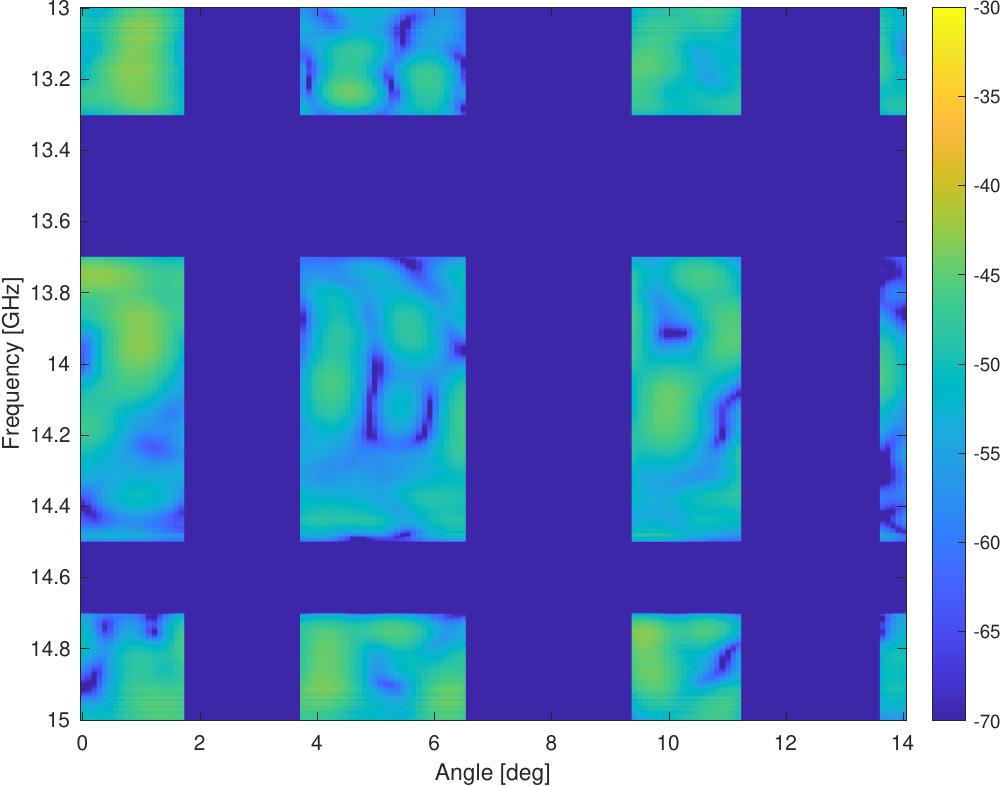}   
			
			\subcaption{}\end{minipage}\hfill 
		\begin{minipage}[b]{0.241\linewidth}   
			\centering   
			\includegraphics[trim=0 0 0 10,clip,width=\linewidth]{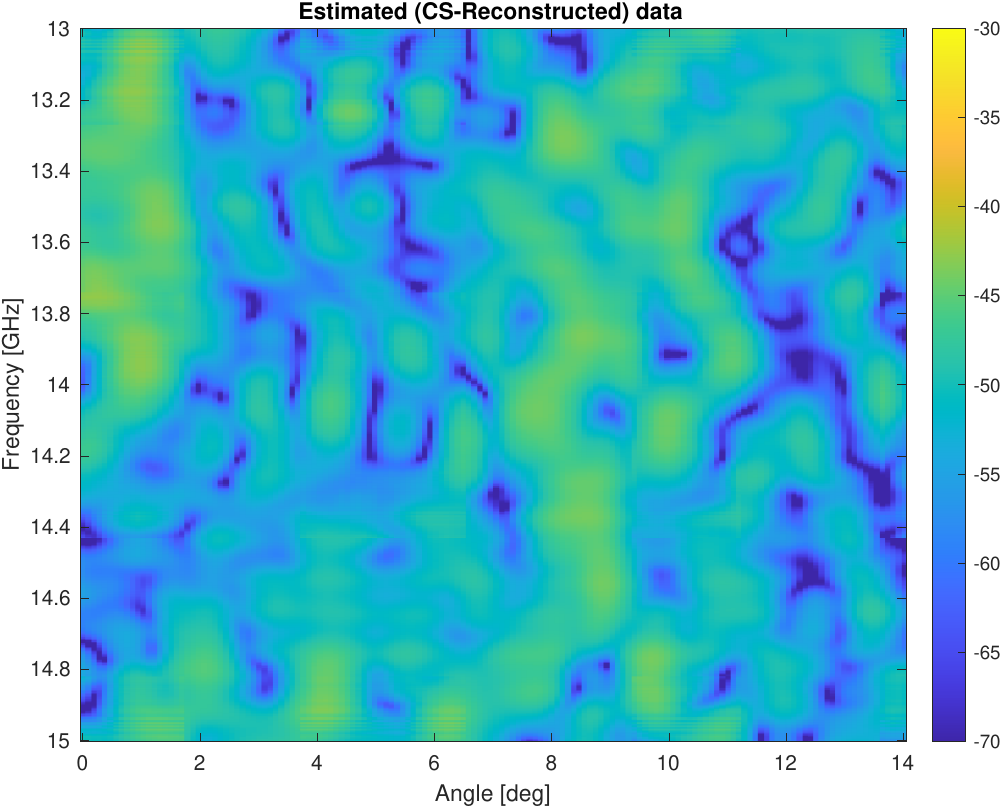}   
			\subcaption{}\end{minipage}\hfill \caption{Frequency-angle (dB) spectrum for a multifunction radar in the presence of two interferers. (a) GT (b) {standard case} (c) {notched case} (d) {N-CS/N-RM} case.}
		\label{fig:map_multi_tgt}
	\end{figure*}

\begin{figure*}[h!]
		\centering  
		\begin{minipage}[b]{0.241\linewidth}   
			\centering   
			\includegraphics[width=\linewidth]{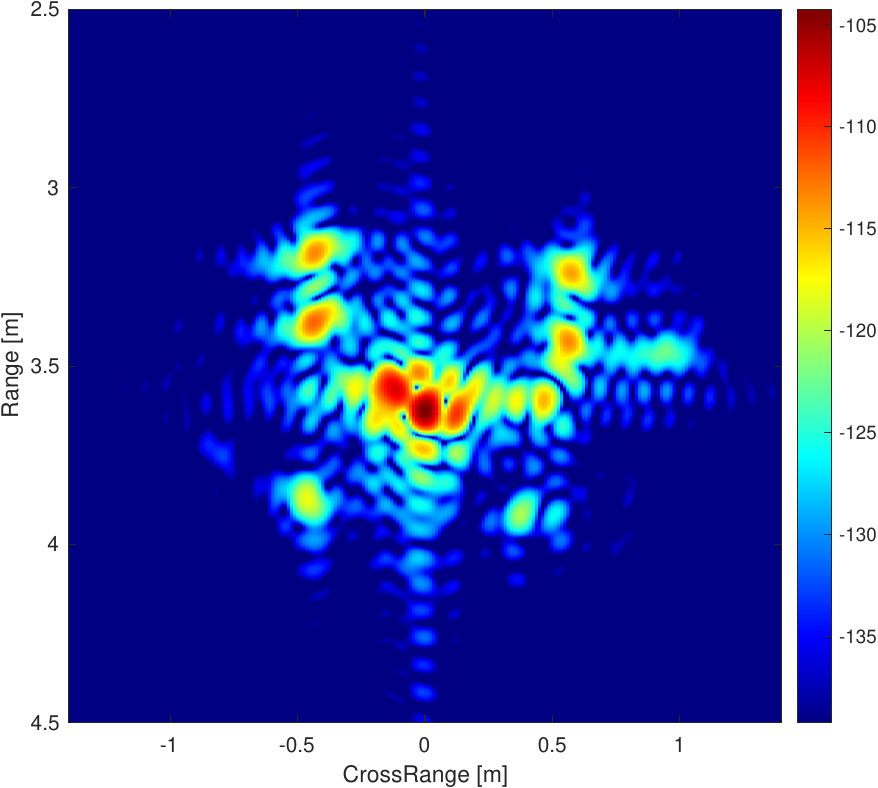}   
			
			\subcaption{}\end{minipage}\hfill 
		\begin{minipage}[b]{0.241\linewidth}   
			\centering   
			\includegraphics[width=\linewidth]{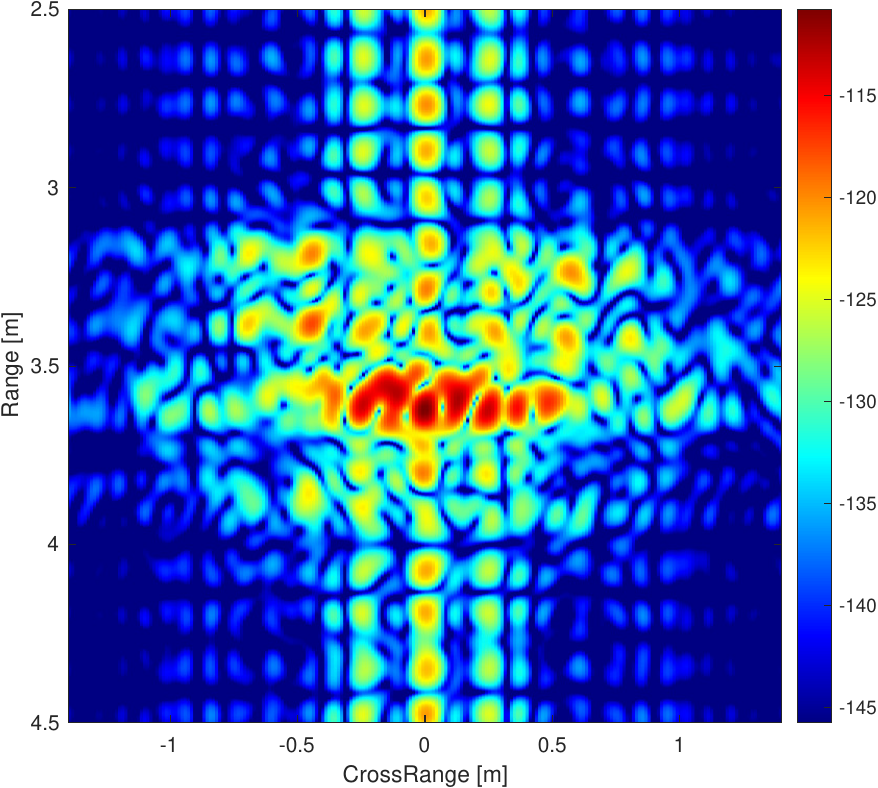}   
			
			\subcaption{}\end{minipage}\hfill 
		\begin{minipage}[b]{0.241\linewidth}   
			\centering   
			\includegraphics[width=\linewidth]{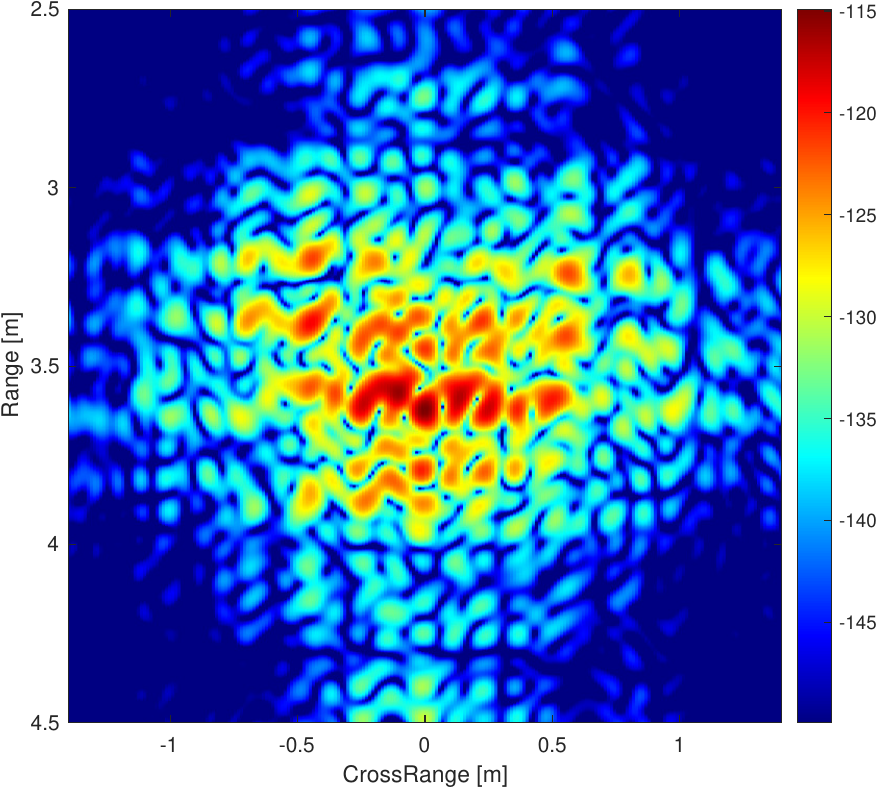}   
			
			\subcaption{}\end{minipage}\hfill 
		\begin{minipage}[b]{0.241\linewidth}   
			\centering   
			\includegraphics[width=\linewidth]{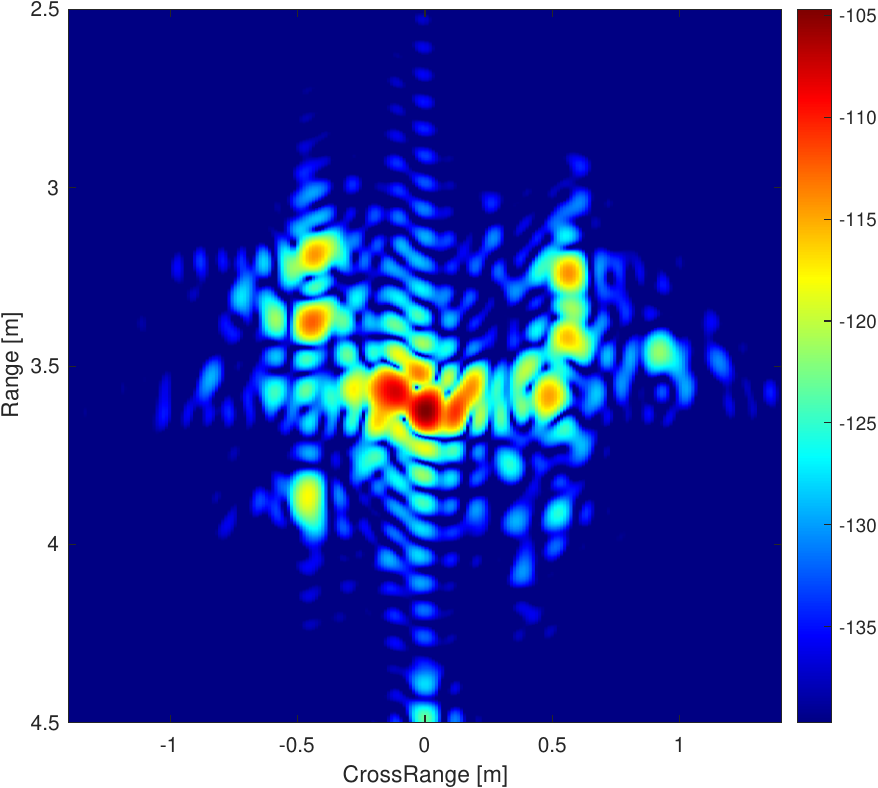}   
			
			\subcaption{}\end{minipage}\hfill \caption{{ISAR image (dB)} for a multifunction radar in the presence of two interferers. (a) GT (b) {standard case} (c) {notched case} (d) {N-CS/N-RM} case.}
		\label{fig:ISAR_multi_tgt}
	\end{figure*}

	The third scenario (Figs.~\ref{fig:map_multi_tgt} and \ref{fig:ISAR_multi_tgt}) considers a multifunction radar that performs a number of {tasks}, including search, track, possible imaging of multiple targets, as well as communication activities~\cite{10412152, 10510313}. Consequently, it is assumed that only 50\% of the dwell time can be used for the ISAR imaging of a target, resulting in the loss of data in the slow-time dimension, in addition to those missing in the frequency domain due to the spectral notches of the cognitive waveform.
	This scenario further corroborates the effectiveness of the considered approach in yielding high-quality ISAR {products} in a contested and congested electromagnetic environment, where the use of a spectrally-shaped waveform is crucial. Furthermore, it demonstrates the potential for radar to perform additional RF operations, thereby encompassing the ISAR process within the framework of a MPAR system.
	
	Table~\ref{tab:extended} reports the {IC, COH, and NMSE values} computed on the obtained ISAR images in all the analyzed scenarios. Inspection of the results {highlights} that the {N-CS}/{N-RM} {framework} attains the highest {score for} all the considered metrics, with values approaching {the corresponding benchmarks}. {This behavior confirms} the capabilities of the considered data recovery methods.
	
	\subsection{Low signal-to-noise ratio (SNR) Scenarios}

	In the following, {the cognitive ISAR process} is investigated in a low SNR regime, such as {in the case of distant targets (still detectable)} as well as in the presence of {some} barrage jamming affecting the entire radar bandwidth. In such contexts, the observed data $\bm{S}$ is significantly corrupted by {interference}.

\begin{figure*}[h]
	\centering  
	\begin{minipage}[b]{0.241\linewidth}   
		\centering   
		\includegraphics[width=\linewidth]{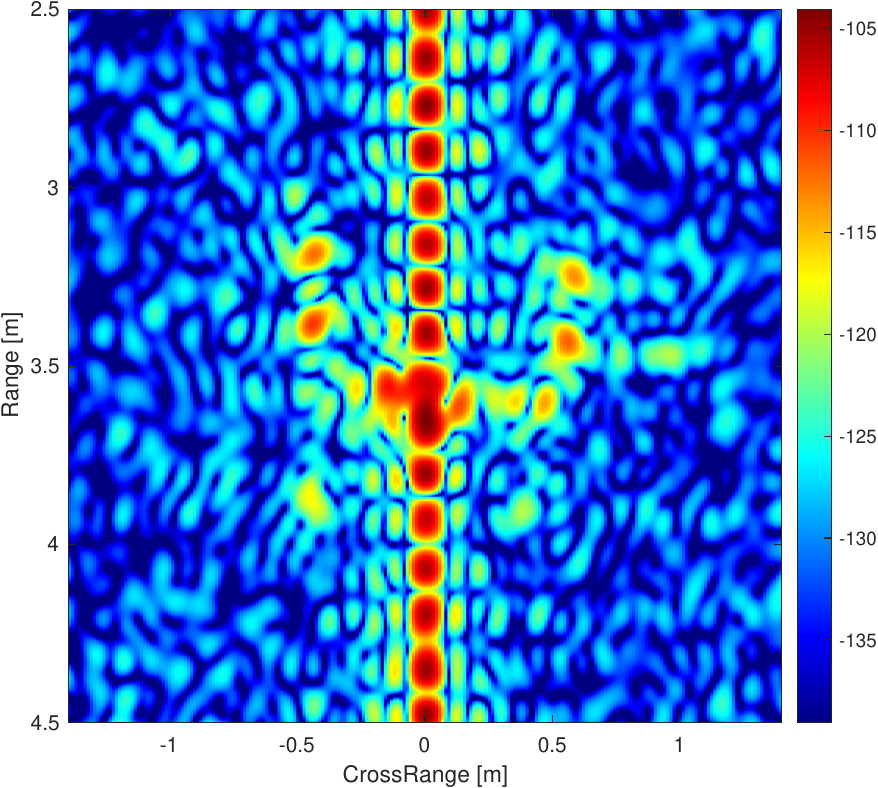}   
		
		\subcaption{}\end{minipage}\hfill 
	\begin{minipage}[b]{0.241\linewidth}   
		\centering   
		\includegraphics[width=\linewidth]{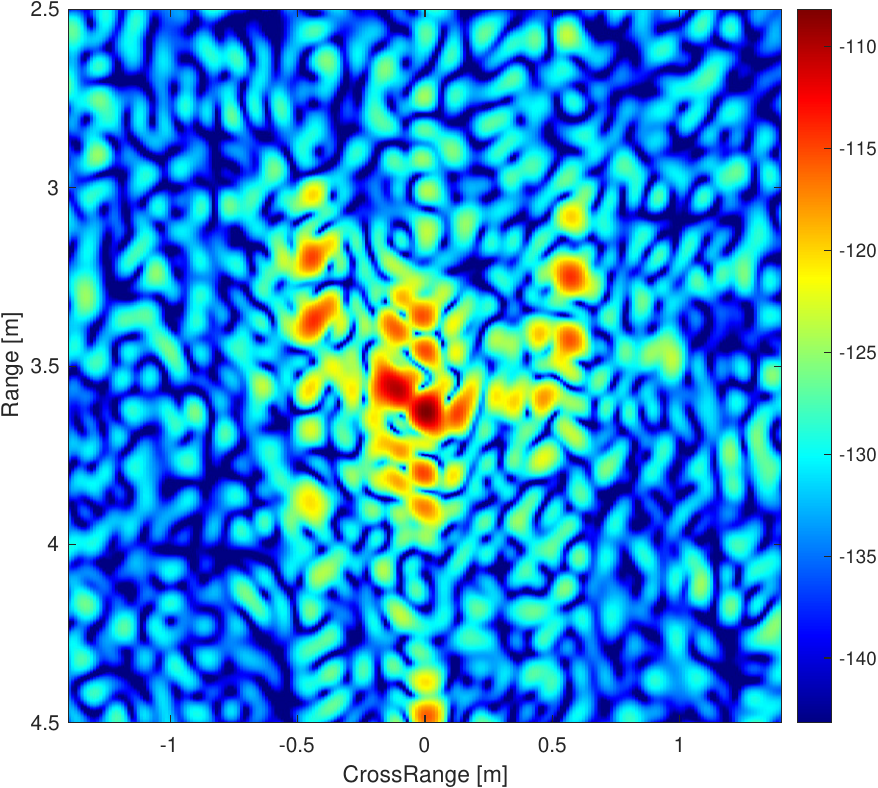}   
		
		\subcaption{}\end{minipage}\hfill 
	\begin{minipage}[b]{0.241\linewidth}   
		\centering   
		\includegraphics[width=\linewidth]{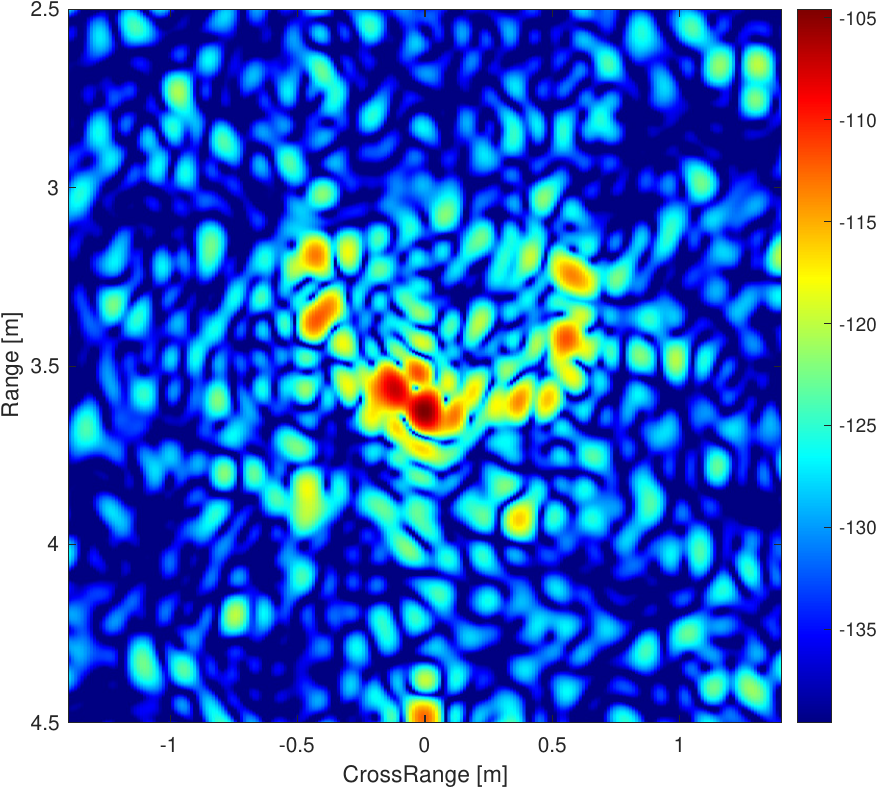}   
		
		\subcaption{}\end{minipage}\hfill 
	\begin{minipage}[b]{0.241\linewidth}   
		\centering   
		\includegraphics[width=\linewidth]{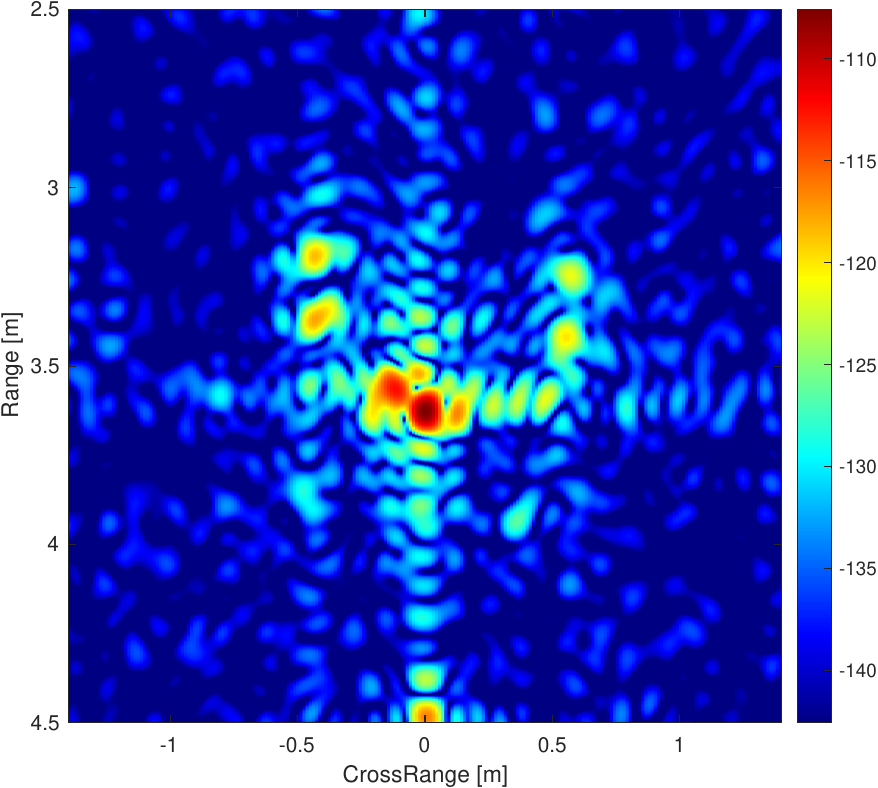}   
		
		\subcaption{}\end{minipage}\hfill \caption{{ISAR image (dB)} in the presence of two interferers and SNR $= -7$ dB. (a) {standard case} (b) {notched case} (c) {N-CS} case (d) {N-RM} case.}
	\label{fig:ISAR_7dB}
\end{figure*}

	Fig.~\ref{fig:ISAR_7dB} reports the ISAR images obtained under various approaches, assuming a single-pulse SNR {(after compression)} of $-7$ dB. The results show that, in both the standard and notched cases, the target signal is overwhelmed by the disturbance, resulting in practically unusable images. Even the CS-based ISAR method, while capable of recovering missing data induced by spectral notches, fails to produce a high-quality image, as evidenced by the presence of spurious scatterers.
	
	In contrast, the proposed N-RM approach demonstrates a strong ability to suppress interference and yield a faithful image of the target. This performance stems from the adopted rank-minimization framework, which enables the reconstruction of slow-time/frequency data by leveraging {only} the dominant modes in the scene.
	
\begin{figure}[t]
	\centering
	\includegraphics[width=0.8\linewidth]{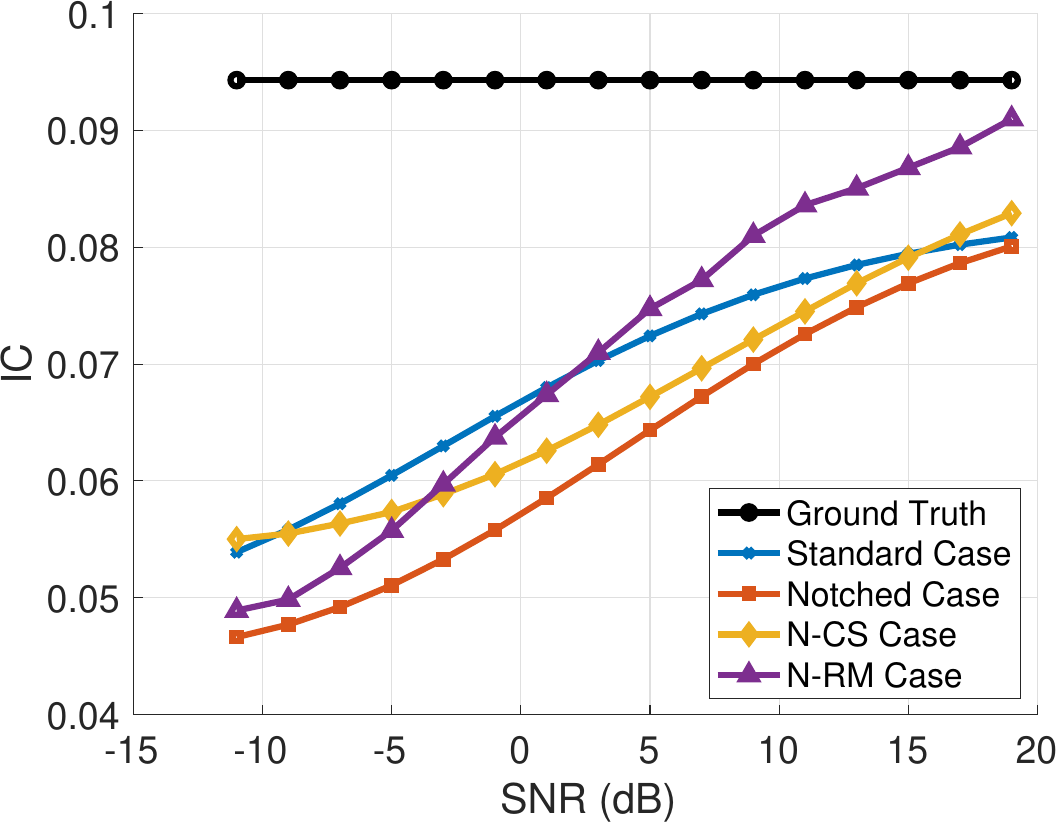}
	\caption{IC vs SNR in the presence of two interferers.}
	\label{fig:IC_vs_SNR}
\end{figure}

\begin{figure}[t]
	\centering
	\includegraphics[width=0.8\linewidth]{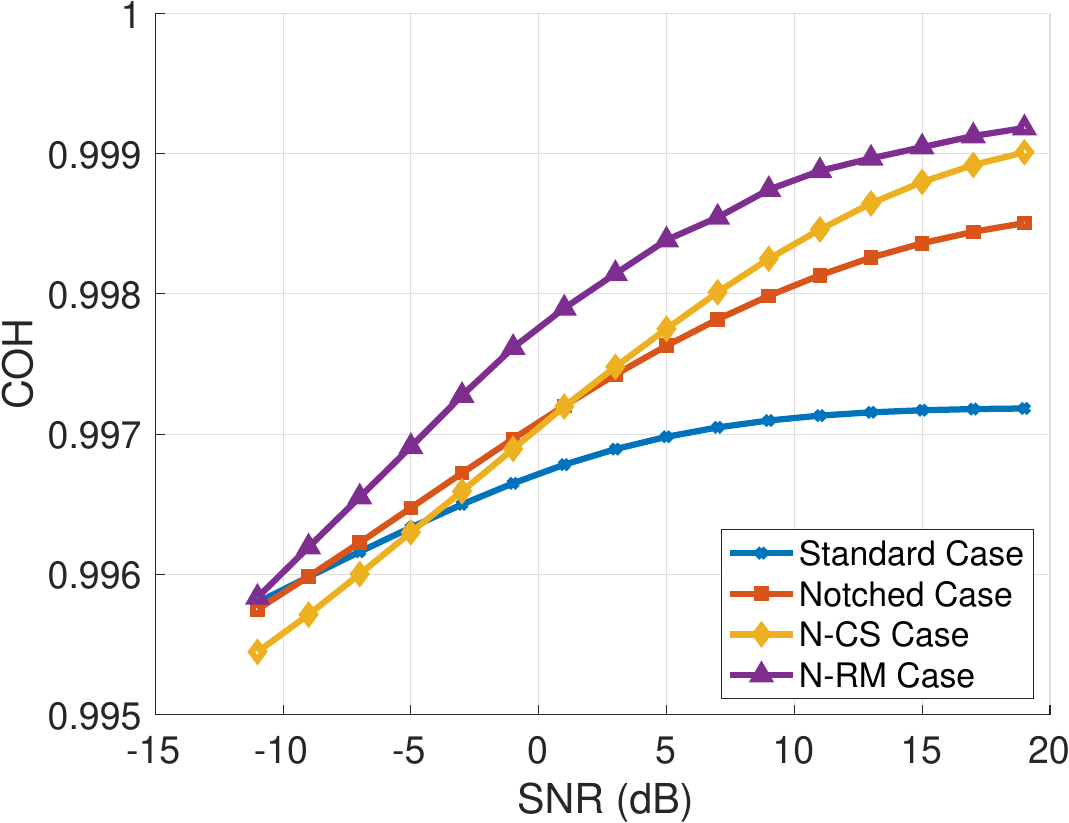}
	\caption{COH vs SNR in the presence of two interferers.}
	\label{fig:COH_vs_SNR}
\end{figure}

\begin{figure}[t]
	\centering
	\includegraphics[width=0.8\linewidth]{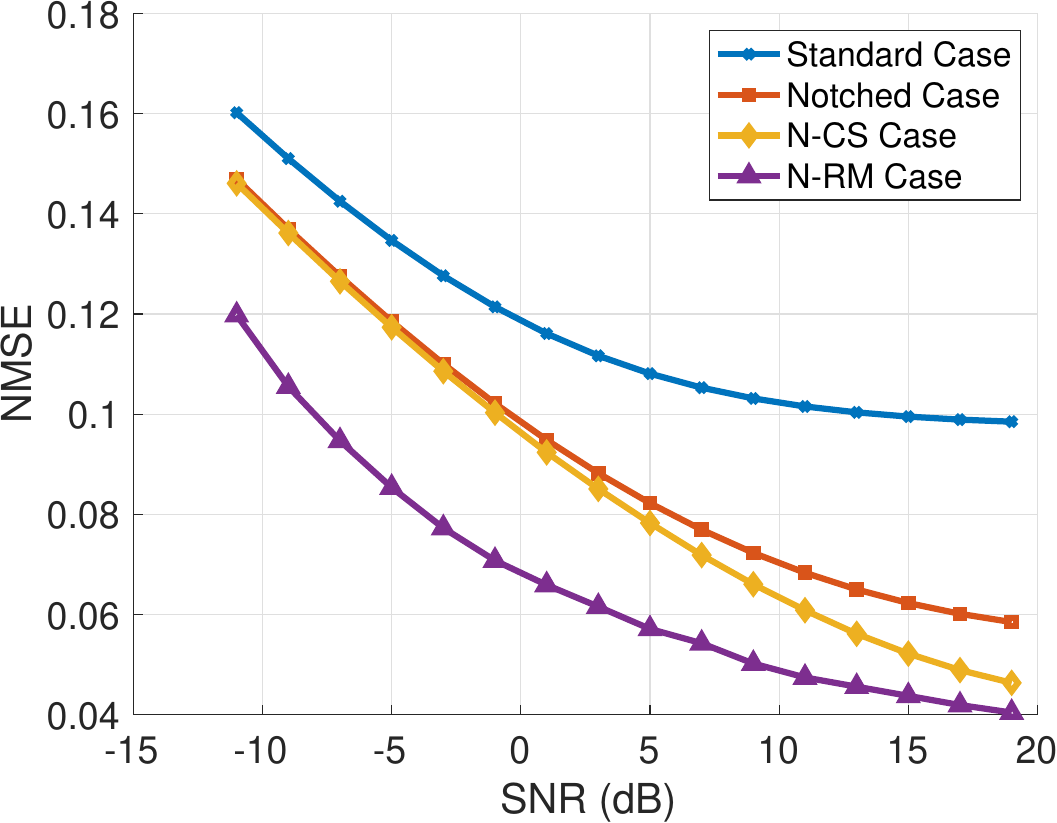}
	\caption{NMSE vs SNR in the presence of two interferers.}
	\label{fig:NMSE_vs_SNR}
\end{figure}

	Furthermore, the image quality, in terms of IC, COH, and NMSE metrics achieved by the {diverse} approaches is illustrated in Figs.~\ref{fig:IC_vs_SNR}, \ref{fig:COH_vs_SNR}, and \ref{fig:NMSE_vs_SNR}, respectively, for SNR values ranging from –11 dB to 19 dB.
	
	{Inspection of the curves highlights} that the proposed N-RM technique consistently outperforms {its} counterparts in terms of image fidelity, as reflected by the higher COH and {lower} NMSE values attained across the entire SNR range. {As compared} to IC, N-RM achieves the highest performance for SNR {greater than} 5 dB, while at lower values, in the standard and N-CS cases, a slightly higher contrast is obtained. However, it is worth remarking that IC solely measures image sharpness and does not reflect fidelity to the ground truth.
	Overall, the proposed missing-data retrieval strategies exhibit {a} robust performance in a variety of practical scenarios, with the N-RM {technique} being particularly effective in challenging low-SNR conditions.
	
	\section{Conclusion}

\begin{table*}[t]
	\caption{Image quality metrics for different scenarios and techniques.}
	\centering
	\label{tab:extended}
	\begin{tabular}{lcc S[round-precision = 4] c}
		\hline\hline
		\textbf{Scenario} & \textbf{Case} & \textbf{IC} & \textbf{COH} & \textbf{NMSE} \\
		\hline
		\multirow{4}{*}{Two interferers} & GT & \num{0.0944} & 1 & \num{0} \\
		& Standard case & \num{0.0839} & \num{0.9986} & \num{0.0600} \\
		& {notched case} & \num{0.0855} & \num{0.9987} & \num{0.0533} \\
		& {N-CS}/{N-RM} case & \bfnum{0.0933} & \bfnum{0.9994} & \bfnum{0.0342} \\
		\hline
		\multirow{4}{*}{{Multiple interferers active in different temporal slots}} & GT & \num{0.0944} & 1 & \num{0} \\
		& Standard case & \num{0.0776} & \num{0.9980} & \num{0.0826} \\
		& {notched case} & \num{0.0824} & \num{0.9988} & \num{0.0540} \\
		& {N-CS}/{N-RM} & \bfnum{0.0898} & \bfnum{0.9994} & \bfnum{0.0348} \\
		\hline
		\multirow{4}{*}{MPAR in the presence of two interferers} & GT & \num{0.0944} & 1 & \num{0} \\
		& Standard case & \num{0.0672} & \num{0.9976} & \num{0.0864} \\
		& {notched case} & \num{0.0731} & \num{0.9982} & \num{0.0647} \\
		& {N-CS}/{N-RM} & \bfnum{0.0900} & \bfnum{0.9990} & \bfnum{0.0451} \\
		\hline
	\end{tabular}
\end{table*}

	In this paper, a cognitive ISAR system designed to guarantee spectral compatibility with possible overlaid emitters has been proposed. To this end, the radar alternates between perception and action stages, exploiting a {spectrum sensing unit} to recognize potential emitters within its frequency range and a tailored waveform design process to synthesize radar imaging waveforms with spectral notches used to probe the environment. A crucial aspect of the proposed architecture is the ability to yield high quality ISAR images, by mitigating the effects of missing data both in the frequency domain, due to spectral notches in the transmitted waveform, and possibly in the slow-time dimension, due to the interrupted mode induced by higher priority tasks in a MPAR system. Therefore, a data recovery process is performed prior to the image formation step, either via the CS framework or by solving a RM problem.
	Extensive numerical analyses, based on drone measurements in the $[13,15]$ GHz band, have {shown} that the proposed system is capable of ensuring spectral compatibility while delivering high-quality ISAR images. Several operating scenarios, including multiple communication sources, multiple emitters active in different temporal slots, as well as a multifunction radar operation, have been considered. The experimental results have shown that the developed strategies allow to produce faithful ISAR images that are visually comparable to the ground truth and effectively maximize standard image quality metrics in several challenging scenarios of practical relevance. Notably, the N-RM method stands out for its high reconstruction accuracy and robustness in challenging scenarios, making it a promising enabler for advanced cognitive-based applications in both civilian and military domains.
	
	Future research studies might involve the application and validation of the proposed technique on additional datasets, as well as a detailed performance analysis under a non-ideal perception stage, where the estimated sources spectral parameters may be inaccurate. Moreover, it will be worth extending the framework to different architectures, such as multistatic and/or polarimetric ISAR systems. Finally, it would be of interest the generalization of the RM recovery process employing the Schatten norm in lieu of the plain nuclear norm.

\section*{Data Availability Statement}
Data supporting this study cannot be made available due to commercial restrictions.
	
	\bibliographystyle{IEEEtran}
	\bibliography{refs}
\end{document}